\documentclass[12pt,preprint]{aastex}

\newcommand{\dmm}{\mbox{$\Delta m_{15}(B)$}}

\def\mbar{\ifmmode\overline{m}\else$\overline{m}$\fi}
\def\Mbar{\ifmmode\overline{M}\else$\overline{M}$\fi}
\def\mibar{\ifmmode\overline{m}_I\else$\overline{m}_I$\fi}
\def\MIbar{\ifmmode\overline{M}_I\else$\overline{M}_I$\fi}
\def\Nbar{\ifmmode\overline{N}\else$\overline{N}$\fi}

\def\ho{\ifmmode H_0\else$H_0$\fi}

\def\dmod{\ifmmode(m{-}M)_0\else$(m{-}M)_0$\fi}
\def\mM{\ifmmode(m{-}M)_0\else$(m{-}M)_0$\fi}
\def\vi{\ifmmode(V{-}I)\else$(V{-}I)$\fi}
\def\viz{\ifmmode(V{-}I)_0\else$(V{-}I)_0$\fi}
\def\EBV{\ifmmode E_{B-V}\else$E_{B-V}$\fi}

\def\kms{\hbox{$\,$km$\,$s$^{-1}$}}

\shorttitle{HST Observations of ESSENCE Supernovae}
\shortauthors{Krisciunas et al.}

\begin{document}
\received{29 April 2005}
\title{Hubble Space Telescope Observations of Nine High-Redshift 
ESSENCE Supernovae\altaffilmark{1,2,3}}

\author{Kevin Krisciunas,\altaffilmark{4}
Peter M. Garnavich,\altaffilmark{4}
Peter Challis,\altaffilmark{5}
Jose Luis Prieto,\altaffilmark{6}
Adam G. Riess,\altaffilmark{7}
Brian Barris,\altaffilmark{8}
Claudio Aguilera,\altaffilmark{9}
Andrew C. Becker,\altaffilmark{10}
Stephane Blondin,\altaffilmark{11}
Ryan Chornock,\altaffilmark{12}
Alejandro Clocchiatti,\altaffilmark{13}
Ricardo Covarrubias,\altaffilmark{10}
Alexei V. Filippenko,\altaffilmark{12}
Ryan J. Foley,\altaffilmark{12}
Malcolm Hicken,\altaffilmark{5}
Saurabh Jha,\altaffilmark{12}
Robert P. Kirshner,\altaffilmark{5}
Bruno Leibundgut,\altaffilmark{11}
Weidong Li,\altaffilmark{12}
Thomas Matheson,\altaffilmark{14}
Anthony Miceli,\altaffilmark{15}
Gajus Miknaitis,\altaffilmark{15}
Armin Rest,\altaffilmark{9}
Maria Elena Salvo,\altaffilmark{16}
Brian P. Schmidt,\altaffilmark{16}
R. Chris Smith,\altaffilmark{9}
Jesper Sollerman,\altaffilmark{17}
Jason Spyromilio,\altaffilmark{11}
Christopher W. Stubbs,\altaffilmark{18}
Nicholas B. Suntzeff,\altaffilmark{9}
John L. Tonry\altaffilmark{8}
and W. Michael Wood-Vasey\altaffilmark{5}
}

%

\altaffiltext{1}{Based in part on observations
with the NASA/ESA \emph{Hubble
Space Telescope,} obtained at the Space Telescope Science Institute,
which is operated by the Association of Universities for Research in
Astronomy, Inc. (AURA) under NASA contract NAS 5-26555. This
research is associated with proposal GO-9860.}
\altaffiltext{2}{Based in part on observations taken at the Cerro Tololo 
Inter-American Observatory, National Optical Astronomy Observatory, which 
is operated by the Association of Universities for Research in Astronomy, 
Inc. (AURA) under cooperative agreement with the National Science Foundation.}
\altaffiltext{3}{Based in part on observations taken with the Very Large
Telescope under ESO program 170.A-0519.}
\altaffiltext{4}{University of Notre Dame, Department of Physics, 225
  Nieuwland Science Hall, Notre Dame, IN 46556-5670;
  {kkrisciu@nd.edu}, {pgarnavi@nd.edu}}
\altaffiltext{5}{Harvard-Smithsonian Center for Astrophysics, 60
  Garden Street, Cambridge, MA 02138; {pchallis@cfa.harvard.edu},
  {mhicken@cfa.harvard.edu}, {kirshner@cfa.harvard.edu},
  {wmwood-vasey@cfa.harvard.edu}}
\altaffiltext{6}{Ohio State University, Department of Astronomy,
  4055 McPherson Laboratory, 140 W. 18th Ave., Columbus, Ohio 43210;
  {prieto@astronomy.ohio-state.edu}}
\altaffiltext{7}{Space Telescope Science Institute, 3700 San Martin Drive,
  Baltimore, MD 21218; {ariess@stsci.edu}}
\altaffiltext{8}{Institute for Astronomy, University of Hawaii,
  2680 Woodlawn Drive, Honolulu, HI 96822; {barris@ifa.hawaii.edu},
  {jt@ifa.hawaii.edu}}
\altaffiltext{9}{Cerro Tololo Inter-American Observatory, Casilla
  603, La Serena, Chile; {caguilera@ctio.noao.edu}, {arest@noao.edu},
  {csmith@noao.edu}, {nsuntzeff@noao.edu}}
\altaffiltext{10}{University of Washington, Department of Astronomy,
  Box 351580, Seattle, WA 98195-1580; {becker@astro.washington.edu},
  {ricardo@astro.washington.edu}}
\altaffiltext{11}{European Southern Observatory, Karl-Schwarzschild-Strasse
  2, Garching, D-85748, Germany; {sblondin@eso.org},
  {bleibund@eso.org}, {jspyromi@eso.org}}
\altaffiltext{12}{University of California, Department of
  Astronomy, 601 Campbell Hall, Berkeley, CA 94720-3411;
  {chornock@astro.berkeley.edu}, {alex@astro.berkeley.edu}, {rfoley@astro.berkeley.edu},
  {sjha@astro.berkeley.edu}, {weidong@astro.berkeley.edu}}
\altaffiltext{13}{Pontificia Universidad Cat\'{o}lica de Chile,
  Departamento de Astronom\'{i}a y Astrof\'{i}sica,
  Casilla 306, Santiago 22, Chile; {aclocchi@astro.puc.cl}}
\altaffiltext{14}{National Optical Astronomy Observatory, 950 N. Cherry Ave.,
  Tucson, AZ 85719; {matheson@noao.edu}}
\altaffiltext{15}{University of Washington, Department of Physics,
  Box 351560, Seattle, WA 98195-1560; {amiceli@astro.washington.edu}, 
  {gm@u.washington.edu}}
\altaffiltext{16}{The Research School of Astronomy and Astrophysics,
  The Australian National University, Mount Stromlo and Siding Spring
  Observatories, via Cotter Rd, Weston Creek PO 2611, Australia;
  {salvo@mso.anu.edu.au}, {brian@mso.anu.edu.au}}
\altaffiltext{17}{Stockholm Observatory, AlbaNova, SE-106 91 Stockholm,
  Sweden; {jesper@astro.su.se}}
\altaffiltext{18}{Department of Physics and Department of Astronomy,
  17 Oxford Street, Harvard University, Cambridge MA 02138; {cstubbs@fas.harvard.edu}}

\begin{abstract}

We present broad-band light curves of nine supernovae ranging in redshift from
0.5 to 0.8. The supernovae were discovered as part of the ESSENCE project, and
the light curves are a combination of Cerro Tololo 4-m and {\it Hubble Space
Telescope} ($HST$) photometry. On the basis of spectra and/or light-curve
fitting, eight of these objects are definitely Type Ia supernovae, while the
classification of one is problematic.  The ESSENCE project is a five-year
endeavor to discover about 200 high-redshift Type Ia supernovae, with the goal
of tightly constraining the time average of the equation-of-state parameter [$w
= p/(\rho c^2)$] of the ``dark energy.'' To help minimize our systematic
errors, all of our ground-based photometry is obtained with the same telescope
and instrument. In 2003 the highest-redshift subset of ESSENCE supernovae was
selected for detailed study with {\it HST}. Here we present the first
photometric results of the survey. We find that all but one of the ESSENCE SNe
have slowly declining light curves, and the sample is not representative of the
low-redshift set of ESSENCE Type Ia supernovae.  This is unlikely to be a sign
of evolution in the population. We attribute the decline-rate distribution of
{\it HST} events to a selection bias at the high-redshift edge of our sample
and find that such a bias will infect other magnitude-limited SN~Ia searches
unless appropriate precautions are taken.

\end{abstract} 

\section{Introduction}

Type Ia supernovae (SNe~Ia) are outstanding probes for cosmology
\citep[see][for extensive reviews]{Fil04, Fil05}. \citet{Ham_etal96b} and
\citet{Rie_etal96} demonstrated that precise values of the Hubble constant can
be obtained using SNe~Ia as distance calibrators, and that Hubble's law is
linear to a high degree of accuracy at small redshifts. \citet{Gar_etal98a}
showed, also using a small sample of high-redshift SNe~Ia, that the matter
density of the Universe must be considerably less than the critical value in an
Einstein-de Sitter universe. \citet{Rie_etal98} and \citet{Per_etal99} found,
surprisingly, that SNe~Ia at high redshift are systematically ``too distant''
(by $\sim$0.25 mag in distance modulus at a redshift of 0.5 compared to a model
of the Universe with $\Omega_M$ = 0.3, $\Omega_{\Lambda}$ = 0.0), implying that
the Universe is expanding at a progressively greater rate.

This deduction is straightforward to understand.  Let us first consider the 
general form of the ``effective'' distance\footnote[19]{\citet[Eq. 20]{Car_etal92}
refer to this as the ``proper motion distance.''  A derivation of Eq. 1
above is given by \citet[Appendix C]{Kri93}.} 
of a galaxy (in megaparsecs) \citep[e.g., ][Eq. 15.49]{Lon84}:

\begin{equation}
d_{\mathrm {eff}} \; = \; \frac{2c}{H_0 \Omega_M^2 (1+z)} \left\{ (\Omega_M -2) 
\left[ (1 + \Omega_M z)^{\frac{1}{2}} - 1 \right] + \Omega_M z \right\} \; ,
\end{equation}

\parindent = 0 mm

where $c$ is the speed of light in km s$^{-1}$, $H_0$ is the Hubble constant 
in km s$^{-1}$ Mpc$^{-1}$, $\Omega_M$ is the mass density of the Universe 
compared to the critical density, and $z$ is the redshift. The relation assumes 
a cosmological constant ($\Lambda$) of zero. For an empty universe this becomes

\begin{equation}
{\mathrm {lim}}_{(\Omega_M \rightarrow 0)} d_{\mathrm {eff}} \; = \; 
\frac{cz}{H_0} \frac  {(1 + \frac{z}{2}) } {(1 + z)} \; .
\end{equation}

\parindent = 9 mm

The observed flux ($F$) of a light source, measured in energy units per
unit area per unit time, is related to the luminosity $L$,
effective distance, and redshift as follows:

\begin{equation}
F \; = \; \frac {L}{4 \pi d_{\mathrm {eff}}^2 (1+z)^2} \; \; \; .
\end{equation}

\parindent = 0 mm

The first factor of (1 + $z$) arises because photons produced at frequency $\nu$
are observed at frequency $\nu$/(1 +$z$); that is, they lose energy due to
the redshift. We need a second factor of (1 + $z$) 
because of time dilation of the arrival of the photons.  


\parindent = 9 mm

We may thus define the ``luminosity distance'' (in Mpc) to be

\begin{equation}
d_{\mathrm {lum}} \; = \; d_{\mathrm {eff}} (1 + z) \; .
\end{equation}

\parindent = 0 mm

The distance modulus is related to the luminosity distance by the standard
equation

\begin{equation}
m - M \; = \; 5 \; {\rm {log}} \; (d_{\mathrm {lum}} \times 10^6)  \; - \; 5 \; = \; 5 \; 
{\rm {log}} \; (d_{\mathrm {lum}} )  \; + \;  25 \; ,
\end{equation}

where the factor of 10$^6$ is used because cosmological distance
is commonly measured in Mpc, not pc.

\parindent = 9 mm

For SNe, we determine the extinction along their lines of sight and 
their extinction-corrected rest-frame apparent magnitudes ($m$) at 
maximum brightness, and then deduce from the light curves the
absolute magnitudes ($M$) at maximum.  The calibration of the absolute 
magnitudes is anchored using nearby SNe~Ia whose luminosities and
distances are consistent with the value of the Hubble constant used
above.\footnote[20]{In point of fact, we determine the distance
{\em ratio} of a high-redshift SN to the low-redshift limit of
the nearby sample.  This effectively eliminates the need to know the 
Hubble constant and the absolute magnitude of a typical SN~Ia in
the analysis.}

In units of $c/H_0$, an Einstein-de Sitter universe 
($\Omega_M = 1, \Omega_{\Lambda} = 0$) 
gives luminosity distances which increase as 
$z + \frac{1}{4}z^2$ over the redshift range of the ESSENCE survey.
In the empty-universe model the
luminosity distances increase as $z + \frac{1}{2}z^2$.  In a
universe with $\Omega_M \approx 0, \Omega_{\Lambda} \approx 1$ the luminosity
distances increase approximately as $z + z^2$.  Note the progression
of the coefficients of $z^2$ and the order of the consequent luminosity
distances.  The point is that, depending on the cosmological parameters,
the loci fan out in the Hubble diagram, and if the observed distance
moduli are larger than one obtains with the empty-universe model, one
must consider a positive cosmological constant.

In practice, we plot the differential
distance moduli (i.e., observed values minus those for an empty-universe
model) vs. the redshift.  Distance moduli of SNe up to $z \approx$ 1.2 are
observed to be systematically larger than we would expect to obtain in 
an empty-universe model. 
\citet[Fig. 6]{Rie_etal04} find that SNe~Ia have distance modulus
differentials which peak at a redshift of 0.46 $\pm$ 0.13. The simplest
deduction is that the Universe must contain matter and some form of
``dark energy''  with a significant negative pressure.  The dark energy behaves
like a non-zero vacuum energy and causes an acceleration of the expansion.

Grey dust along the line of sight \citep[e.g.,][]{Agu99} or SN~Ia
evolution in average luminosity over several billion years could explain
the implied faintness of SNe at $z \sim 0.5$. Another concern would be
selection effects in the discovery of high-redshift SNe.  
\citet{Li_etal01} and \citet{Ben_etal05} discuss the diversity of SNe~Ia
and how the relative numbers of the different sub-types are affected in
magnitude limited surveys.  \citet{Clo_etal00} and \citet{Ho05} discuss
the effect of contamination in cosmological surveys by Type Ibc SNe.  
Examples such as SN~1992ar were even brighter than the brightest SNe~Ia at
maximum light, but, on average, stripped core SNe are two magnitudes
fainter than SNe~Ia.  Without high quality spectra or three (or more)
filter photometry \citep[see, for example,][]{Poz_etal02,Gal_etal04} it
might be difficult to distinguish between Types Ia and Ibc.  Since the
ESSENCE project relies primarily on two-band photometry and rather
``grassy'' spectra, our ability to distinguish between SNe~Ia and Type Ibc
SNe is limited; these limitations apply to other surveys too.

Recently, however, \citet{Ton_etal03},
\citet{Kno_etal03}, and \citet{Bar_etal04} have tested supernova
systematics out to $z \approx 1$, confirming and strengthening the 
evidence for an acceleration.  \citet{Rie_etal01} and \citet{Rie_etal04} 
have pushed the boundary for SN discoveries out to a redshift of $z \approx$ 1.7
using the {\it Hubble Space Telescope} ($HST$).  At $z \gtrsim 1.2$ the SNe 
are {\em brighter} on average than one would expect compared to the 
empty-universe model. On this basis we can reject the notion of a 
significant amount of grey dust along the line of sight, as its effect 
would presumably be even greater at higher redshifts. At $z \gtrsim
1.2$ we observe the Universe when it was small enough that the
gravitational attraction of all the matter exceeded the
repulsive effective of the dark energy; the Universe was therefore 
{\em decelerating}. Indeed, Riess et al. (2004) find that
the transition from deceleration to acceleration could have occurred 
as recently as $z \approx 0.5$.

A key part of the present-day concordance model of the Universe is some
form of dark energy.  If the equation-of-state parameter $w \equiv
p/(\rho c^2$) = $-$1, where $p$ is the pressure and $\rho$ is the
density, then the dark energy is in the form of a standard
cosmological constant \citep[e.g., ][]{PeebRat03}.

Dark-energy models with $w \neq -1$ require internal degrees of
freedom, or the presence of non-adiabatic stress perturbations, to
remain gravitationally stable \citep{Hu05}.  If $w > -1$, the dark-energy 
density slowly decreases as the Universe expands;  $w \geq -1$
is required by the ``weak energy condition'' of general relativity
\citep{Gar_etal98b}.  The case of $w < -1$ was discussed by 
\citet{Car04}.  It results in an ever-faster expansion, leading to
a ``Big Rip.''  Such an energy was postulated by \citet{Cal02}
and dubbed a ``phantom field.''

\citet{Upa_etal05} present an excellent summary of the current
constraints and forecasts for dark energy.  They characterize
the form of the equation-of-state parameter as $w = w_0 + w_1 z$
out to $z$ = 1 and find that existing data for the cosmic
microwave background, SNe~Ia, and the galaxy power spectrum give
$w_0 = -1.38^{+0.30}_{-0.55}$ (2$\sigma$), and $w_1 =
1.2^{+0.64}_{-1.06}$ (2$\sigma$).  After discussing ongoing and
planned experiments, they conclude that, ``unless we are lucky
enough to find a dark energy that is very different from the
cosmological constant [$w \equiv -1$], new kinds of measurements
or an experiment more sophisticated than those yet conceived
will be needed in order to settle the dark-energy issue.''

What is the mass density of the Universe?  \citet{Eis_etal05} 
investigate the large-scale structure of the Universe using nearly
47,000 luminous red galaxies from the Sloan Digital Sky Survey (SDSS).
They find that $\Omega_M$ = 0.273 $\pm$ 0.025 + 0.123(1+$w_0$) + 
0.137$\Omega_K$, and that the curvature term $\Omega_K$ is 
statistically equal to zero ($-$0.010 $\pm$ 0.009).  
However, \citet{Col_etal05} derive a lower value
of the mass density; their revised result from over 220,000 galaxy
redshifts in the Two-degree Field (2dF) redshift survey 
is $\Omega_M h$ = 0.168 $\pm$ 0.016.
With $h \equiv H_0/100$ = 0.72 $\pm$ 0.06 from \citet{Fre_etal01},
we obtain $\Omega_M$ = 0.233 $\pm$ 0.030.

The first results from the Wilkinson Microwave Anisotropy Probe (WMAP)
have shown that the spatial geometry of the Universe is flat
\citep{Ben_etal03}; the total energy density of the
Universe corresponds to $\Omega_{tot} \; = \;  \Omega_M \; + \;
\Omega_{\Lambda}$ = 1.02 $\pm$ 0.02.  \citet{Teg_etal04} use data from
WMAP and SDSS to constrain the Hubble constant $H_0$ = 70$^{+4}_{-3}$
km s$^{-1}$ Mpc$^{-1}$, the matter component of the Universe
$\Omega_M$ = 0.30 $\pm$ 0.04, and neutrino masses to less than
0.6 eV. The addition of SNe~Ia to the mix gives an age of the
Universe of $t_0$ = 14.1$^{+1.0}_{-0.9}$ Gyr.

Ever since the luminosities at maximum of SNe~Ia were
first convincingly shown to be related to the post-maximum rate of
decline \citep{Phi93}, SNe~Ia have come to be regarded as the most
common reliable cosmic beacons with which to determine extragalactic distances
beyond 20 Mpc.  \citet{Ham_etal96c}, \citet{Phi_etal99},
\citet{Ger_etal04}, and \citet{Pri_etal05} have refined the \dmm\
method, where the decline-rate parameter is defined to be the number of
$B$-band magnitudes that a SN~Ia declines in the first 15 days
after maximum. This method previously used the $BVI$ light curves and
now uses $R$-band photometry as well. The ``stretch method'' 
\citep{Per_etal97, Gol_etal01} scales $B$-band and
$V$-band templates in the time domain to fit actual light curves. The
Multi-color Light-Curve Shape (MLCS) method of \citet{Rie_etal96} and
\citet{Rie_etal98} uses the $BVRI$ light curves to give a parameter
$\Delta$, the number of magnitudes that a SN~Ia is brighter than
($\Delta < 0$) or fainter than ($\Delta > 0$) a fiducial light curve.  
\citet{Jha02} and \citet{Jha_etal05} expand the MLCS method to include
$U$-band data.  This is very important for studies of high-redshift
SNe, because photometry of objects at $z > 0.8$ observed in the
$R$ band corresponds to rest-frame photometry in the $U$ band. Finally,
\citet{Wan_etal03} have shown that useful information such as reddening
can be derived from plots of the filter-by-filter magnitudes vs. the
photometric {\em colors} (instead of vs. time), using data from the
month after maximum light.



The ESSENCE project has as its prime objective the measurement of the
time average of the equation-of-state parameter $w$ to an accuracy of 
$\pm$10\%.  A description of the strategy and methodology of the project is
given by \citet{Mik_etal05}.  Briefly stated, over a five-year period we
shall discover roughly 200 SNe~Ia at $z = 0.2$--0.8 using
the facility CCD mosaic camera on the 4-m Blanco telescope at Cerro
Tololo Inter-American Observatory (CTIO).  In the past we have observed
individual SNe with up to seven telescope/camera combinations.  By
obtaining all of the ground-based ESSENCE photometry with a single
telescope and camera, we should be able to better control systematic 
photometric errors.

In this paper we present a small fraction of our eventual sample
of 200 SNe~Ia.  Each object presented here was observed with the
CTIO 4-m telescope through the $R$ and $I$ bands, and with $HST$
using the Advanced Camera for Surveys (ACS) and its F625W,
F775W, and F850LP filters.  The nine objects presented here were
members of our highest redshift subsample.  Since one of the
goals of the ESSENCE project is to identify and minimize sources
of systematic error, observing some of our highest redshift
objects with $HST$ had obvious practical advantages since $HST$
gave the highest S/N photometry.

Spectra of the SNe themselves and/or the host galaxies
are being obtained with the two Keck telescopes, the VLT, Gemini North
and South, both Magellan telescopes, the MMT, and at the Fred L. Whipple
Observatory (Mt. Hopkins). \citet{Mat_etal05} describe the
spectroscopic aspects of the SN search and discuss results obtained
thus far using the spectral cross-correlation program SNID
(Tonry et al. 2005, in preparation).  \citet{Sun_etal05} have shown that 
there are insignificant systematic differences ($\sim$0.02 mag)
between $HST$ photometry with WFPC and ground-based photometry.
We assume that $HST$/ACS has been tested and calibrated sufficiently
well that publicly available photometric zeropoints and filter
profiles allow us to combine ACS data and ground-based
photometry without problems.  For one of our objects we also obtained
three orbits of data with the $HST$ infrared camera NICMOS.

\section{Observations}

In the first two seasons of the ESSENCE project we discovered 46
definite SNe~Ia, 6 likely SNe~Ia, 5 core-collapse SNe, plus a number of
candidates that were neither confirmed nor rejected as SNe; see
\citet{Mat_etal05} for a thorough discussion.
In the third year of the ESSENCE project we discovered 45 additional
SNe, of which 30 are definitely SNe~Ia, 10 are possible SNe~Ia, and
5 are core-collapse SNe.

Are these relative numbers sensible?  \citet{Dah_etal04} considered rates of
SNe found in the Great Observatories Origins Deep Survey (GOODS).  These
authors found 17 SNe~Ia to $z$ = 1.0 and 16 core-collapse SNe (Types II and
Ibc) to $z$ = 0.9.  Given that GOODS found SNe~Ia to $z$ = 1.6, we may consider
GOODS close to being volume limited at the lower redshifts just stated.  This
is what is being found with nearby SNe discovered by the Katzman Automatic
Imaging Telescope \citep{Lea_etal04} $-$ comparable numbers of SNe~Ia and
core-collapse SNe are found in a volume limited sample.  ESSENCE, however, is a
{\em magnitude limited} survey.  On average, core collapse SNe are at least 2
mag fainter at maximum compared to SNe~Ia.  If we discover SNe~Ia out to $z
\sim$ 0.8, we discover core-collapse SNe to about $z \sim$ 0.45.  ESSENCE
should be finding between 10 and 20 times as many SNe~Ia compared to
core-collapse SNe on the basis of the relative volumes being sampled.  Since
our experiment has as its goal the discovery of 200 SNe~Ia, we even try to
select against Type II SNe by imposing a color cut on the candidates flagged as
potential objects of interest. Since Type II SNe are very blue prior to maximum
light, we ignore candidates with $R-I < 0$.  This increases the chance that a
spectrum of a candidate will reveal a SN~Ia.  Thus, the low percentage of
core-collapse SNe found in our survey is a measure of our success.

In this paper we report photometry of nine objects discovered in October,
November, and December 2003 which were also observed with $HST$/ACS
(using the Wide Field Camera). The goal of the $HST$ observations was
to observe the highest-redshift end of the ESSENCE sample since
photometry becomes difficult with the CTIO 4-m as they fade.
In Table \ref{obstab} we list the nine
objects.  Their redshifts were obtained from observations with Keck I +
LRIS, Gemini + GMOS, VLT + FORS1, or Magellan (Baade + IMACS).
Spectroscopic details are given by \citet{Mat_etal05}.

Figure \ref{stamps} shows $HST$/ACS images, taken through the F775W filter, 
of the nine SNe discussed in this paper.  Each represents
the combination of two integrations per date and from four to six
dates per object.  The total integration times were 3700~s (e510),
5100~s (SN~2003jo), and 4400~s (the other seven SNe).  As one can see, only
SN~2003jo, SN~2003ll, and SN~2003li were hosted by galaxies sufficiently 
large and bright to show significant structure.  It is not entirely
clear that the galaxy near SN~2003kv is the host of that SN.
Several of our objects were effectively ``hostless'' SNe, meaning that
the hosts, if they exist, have surface brightness too low to be
detected in $HST$ and ground-based images.

The program SNID indicates that SN~2003ku is a SN~Ia,
but there is ambiguity regarding its true redshift.  While the
most likely redshift on the basis of the spectrum is 0.79, the next most
likely redshift is 0.41; see \S~3.3 and \citet{Mat_etal05} 
for more comments.

We observed supernova e510 spectroscopically on 29 November 2003 (UT dates are
used throughout this paper). We obtained one 1800~s spectrum with Gemini 
North + GMOS, but were then shut down by high humidity, and we were unable to 
obtain a redshift from the nearly featureless spectrum. We also attempted 
to get spectra with Keck, but were hampered by high humidity and clouds. 
In mid-October 2004, long after the SN had faded, we attempted to get a
redshift from spectra of the faint host galaxy.  We took three 1800~s
spectra with Gemini North + GMOS, but there was insufficient signal to
extract a usable spectrum which would provide a redshift. However,
following the guidelines of \citet{Rie_etal01} and \citet{Bar_Ton04},
one can derive a redshift-independent distance to a {\em set} of SNe~Ia,
provided one has sufficiently good photometric coverage in more than
a single passband.  For any individual object the results are understandably
somewhat uncertain.  Analysis using the code of \citet{Pri_etal05}
gives a minimum reduced $\chi^2$ value of the light-curve fits at a
redshift of 0.68 for e510.  However, any redshift between 0.64 and 0.84
will give a reduced $\chi^2$ value less than 1.0 for this object.  It
is best if we eliminate it from further consideration.  Because there
is no spectral information for e510, it never received an IAU designation.

\citet{Mat_etal05} indicate that SN~2003kv was probably a SN~Ia, but the
spectroscopic identification was not certain.  Photometric analysis (below)
is entirely consistent with the notion that SN~2003kv was a SN~Ia.


The ESSENCE supernovae were not observed as ``targets of opportunity'' (TOOs)
with $HST$ since these are very disruptive to the scheduling of the telescope.
Instead, a ``pseudo-TOO'' method was employed which was pioneered
by the high-redshift supernova search teams several years ago. We
specify months before observation (in the Phase II submission) the region
on the sky to be searched,  and we specify dates that exact supernova positions
will be available for insertion into the $HST$ schedule. Usually it is
five days to a week before $HST$ begins observing the supernovae after
the coordinates have been delivered.

Generally we have a number of supernovae in each field from which to choose
for $HST$ observations, but there are times when a non-optimal target
must be inserted at the deadline. Also, because of the delays between 
the supernova discovery, spectral confirmation, and the $HST$ schedule 
upload, the supernovae are rarely observed before maximum
light with $HST$. Since $HST$ was meant to improve the quality of photometry
at late times for our faintest targets, this delay is not a major
problem for this study.

Our $HST$/ACS data were obtained using the F625W, F775W, and F850LP
filters, which are essentially the same as the $r^{\prime}$,
$i^{\prime}$, and $z^{\prime}$ filters of the SDSS photometric system
\citep{ASmi_etal02}.  The throughput transmission curves with $HST$/ACS
are shown in the Appendix.  One of our objects was observed in the
near-infrared with NICMOS and its F110W filter (similar to a $J$-band
filter).

$HST/ACS$ photometric calibration is based on the zeropoint values of
\citet{Sir_etal05}, which in turn are based on the Vega spectrophotometric
calibration of \citet {Boh_Gil04}.  For an aperture of 50-pixel radius, the
Vegamag zeropoints are zp$_{F625W}$ = 25.731, zp$_{F775W}$ = 25.256, and
zp$_{F850LP}$ = 24.326 mag \citep[][Table 11]{Sir_etal05}.

Our $HST/ACS$ magnitudes are based on small-aperture photometry (4-pixel 
radius), using aperture corrections to $r$ = 50 pixels consistent with
the radial profiles delineated in Table 3 of \citet{Sir_etal05}.
This aperture photometry was done using the
{\sc apphot} package within IRAF.\footnote[21] {IRAF is distributed by
the National Optical Astronomy Observatory, which is operated by AURA, Inc. 
under cooperative agreement with the National Science Foundation.} 

In the case of SN~2003jo we obtained $HST$ template images on 22
May 2004, some 212 observer-frame days after the date of maximum light
(or 139 rest-frame days).  SN 2003jo was faintly visible in the F625W and
F775W images on that date, but not visible in the F850LP images. To remove
the light of the host galaxy we subtracted the templates, then made
corrections (0.02 to 0.06 mag) to the photometry based on the late-time detections.

Template images of SN~2003lh were obtained on 8 June 2004, some
117 rest-frame days after $t(B_{max})$.  The SN is visible in the F775W
template and undoubtedly is present in the other two.  For $HST$
photometry of this object, we could perform image subtraction to eliminate any
effect of host-galaxy light, then correct for the presence of the SN in
the templates using light curves of a different slow decliner such as
SN~1991T \citep{Schmidt_etal94, Lir_etal98}.  We determined that these
corrections would be as large as $\sim$0.2 mag.  Or, we could simply perform
small-aperture photometry (4-pixel radius) without image subtraction.  We
found that the two methods gave the same final photometric values within
1$\sigma$, and we adopt the photometry from the latter method, as 
it was simpler and relied on fewer assumptions.

$HST$ template images of SN~2003ll were obtained on 5 October 2004.  
The other objects discussed in this paper showed no significant
host-galaxy light near the locations of the SNe in the $HST$ images.

In Table \ref{hstphot} we list the photometry of the SNe using ACS/WFC
with $HST$.  We also give three orbits of F110W data for one SN,
obtained with $HST$ and its near-infrared camera NICMOS.

The ground-based images with the CTIO 4-m telescope were taken through
$R$-band and $I$-band filters, the details of which are described in the
Appendix. Our $R$ filter is essentially the same as the \citet{Bes90} $R$
filter, but our $I$-band filter has steeper short-wavelength and 
long-wavelength cutoffs than Bessell's $I$-band filter.

We found that the ground-based imaging in the $R$ band gave higher
signal-to-noise ratio (S/N) detections for the objects with $z
\lesssim$ 0.6, but that the $I$-band imaging gave higher S/N detections
for the objects with $z \gtrsim$ 0.6. This is just an empirical
consequence of the spectral energy distributions of SNe~Ia in
the rest-frame $UBV$ bands coupled with the quantum efficiency of
the CTIO 4-m mosaic camera in $R$ and $I$.

Ground-based photometry was carried out using template images obtained at
least 18 {\em rest-frame} days prior to the observed maximum, or, in one
case, long after the SN had faded.  (At 18 to 20 rest-frame days prior to
maximum light, any light of a high-$z$ SN on the rise would be lost in
the sky noise of the CTIO 4-m images.) When possible we used a median of
three template images obtained on photometric nights to eliminate cosmic
rays in the templates.  For eight of the nine objects discussed here we
used reference images from early in the 2003 observing season. In the
case of SN~2003jo we used reference images from 28 November 2002 ($R$
band) and 11 October 2004 ($I$ band).

For the rotation, alignment, kernel matching, and difference imaging of
the ground-based images we used two packages of scripts written by one of
us (BPS).  The result is point-spread function (PSF) magnitudes of field
stars and the SNe themselves. The scripts rely on the kernel-matching
algorithm of \citet{Ala_Lup98}.  For reasonably high S/N detections of
the SNe, we adopted the 1$\sigma$ symmetrical error bars in magnitudes
from {\sc dophot}.  For faint signals, the error bars in magnitude space
are not symmetrical.  We then chose a medium-sized number (20) of random
locations in the subtracted images (avoiding obvious image defects) to
derive the sky noise, and derived 1$\sigma$ error bars in flux space,
which we then converted to asymmetrical upper and lower errors in
magnitude space. In the case of the CTIO 4-m $R$-band images, the sky
noise is roughly 250 analog-to-digital units (ADUs, or counts, where 1
ADU $\approx$ 1.8 $e^-$) in 200~s exposures (templates and data images)
on clear, moonless nights.  400~s $I$-band images give corresponding sky
noise of about 450 ADUs in the subtracted images.  A SN~Ia at maximum
light and $z \approx 0.5$ typically gives a signal of 4000 ADUs in
corresponding $R$-band and $I$-band exposures obtained with the CTIO 4-m
telescope.

We used data from the early data release of the SDSS 
to obtain $R$-band and $I$-band magnitudes for the field stars near the SNe
discussed in this paper, relying on the transformations given in
Table 7 of \citet{ASmi_etal02}.  We avoided using field stars with
$r^{\prime} - i^{\prime}$ $>$ 0.95 mag, as many stars this red are variable
and the scatter in the photometric transformation becomes large.








To check our algorithm for calibrating the SN photometry, we obtained images
of six of our nine SN fields on 20 October 2004 using the CTIO 0.9-m
telescope.  This was a photometric night, allowing the determination of the
atmospheric extinction values and the instrumental color terms using
observations of seven \citet{Lan92} fields. From aperture photometry with an
8 pixel ($3.2''$) radius aperture, we obtained sensible photometry to $R$ =
19.7 mag with 600~s to 720~s exposures.  In 900~s $I$-band exposures we
effectively reached magnitude 19.0.  This provided $\sim$1.6 mag of overlap
with the brightest unsaturated stars in our CTIO 4-m images (which were
typically 200~s in $R$ and 400~s in $I$).  A comparison of the derived $RI$
magnitudes and the values obtained from the transformation from SDSS
$g^{\prime}r^{\prime}i^{\prime}$ magnitudes is shown in Figure \ref{RI_test}.  
The $R$-band differentials give a slope of $-$0.0049 $\pm$ 0.0031 mag per
mag, while the $I$-band differentials give a slope of $-$0.0087 $\pm$ 0.0058.  
$\Delta$ magnitude is in the sense ``observed values from CTIO 0.9-m
photometry directly tied to \citet{Lan92} standards'' {\em minus} ``values
derived from SDSS photometry''. Neither slope differs from zero at a
statistically significant level. At $R$ = 19.0 mag, $\langle\Delta R \rangle
= -0.027 \pm 0.051$ mag, while at $I$ = 19.0 mag, $\langle \Delta I \rangle =
-0.025 \pm 0.046$ mag. Neither differs significantly from zero.  We are
trying to account for all sources of systematic error greater than 0.01 mag
in the ESSENCE photometry.  Hence, a more robust test involving more fields
and nights is warranted.

In Table \ref{ctio4mphot} we give the photometry of the SNe using the
CTIO 4-m telescope and its facility mosaic camera.  We list the $R$ and
$I$ magnitudes in the natural magnitude system of the 
telescope/filter/detector transmission function, with a zero point based
on the \citet{Lan92} system.  To force the zero point of the natural system
to that of Landolt, we plot ($R_{nat} - R_{Landolt}$) as a function of
($V-R$) and ($I_{nat} - I_{Landolt}$) as a function of ($V-I$).  By forcing
($R_{nat} - R_{Landolt}$) and ($I_{nat} - I_{Landolt}$) to be 0.00 where
($V-R$) and ($V-I$) are zero, we transfer the zero point of the Landolt
system onto the natural system.  See the Appendix of \citet{Schmidt_etal98}.

In Figures \ref{R_light_curves} and \ref{I_light_curves} we show the
observed light curves of the nine SNe discussed here.  Because the
central wavelengths of the $R$-band and F625W filters are similar, we
would expect that photometry in those bands would be reasonably
similar. Also, the $I$-band photometry should be similar to the F775W
and F850LP photometry. An exception occurs with $R$-band and F625W
photometry if we are observing a SN~Ia with $z \approx$ 0.8;  
the flux of the SN just longward of the Ca II H \& K lines is
included in the $R$-band photometry but excluded from the F625W band.  
This can make a difference of $\sim$0.3 mag three weeks after
maximum light in the observer's frame.

\section{Discussion}

\subsection{Light-Curve Fits and a Composite Spectrum}

Seven of the nine objects discussed in this paper have unambigous redshifts
on the basis of spectra \citep{Mat_etal05}.  We were unable to obtain a
redshift of e510 from the SN itself when it was visible, or from the faint
host galaxy the following year, so it was never assigned an official name by
the IAU.  But its light-curve shapes and maximum magnitudes are completely
compatible with those of other high-redshift SNe~Ia.  The case of SN~2003ku
is discussed below.

In order to derive the distances of the SNe, we first K-corrected the
ground-based and space-based photometry to rest-frame $U$, $B$, $V$, or
$R$ photometric bands. For the MLCS method of \citet[][hereafter
MLCS2k2]{Jha_etal05} and the \dmm\ analysis \citep{Pri_etal05}, if $z \leq
0.6$ the ground-based $R$-band and F625W data were K-corrected to
rest-frame $B$, ground-based $I$-band and F775W data were transformed to
rest-frame $V$, and F850LP data were transformed to rest-frame $R$.  
Photometry of SN~2003kp was transformed to $B$ and $V$. For SN~2003kv 
we transformed the photometry to $UBV$ instead of $BVR$.

In Tables \ref{mlcs_fits}, \ref{batm_fits}, and \ref{lc_fits}, we give
the light-curve fits using MLCS2k2, the
Bayesian Adapted Template Method \citep[BATM;][ and Tonry et al. 2005, in
preparation] {Ton_etal03}, and the \dmm\ method of \citet{Pri_etal05},
respectively.  Our derived distance moduli are consistent within the
errors using the three light-curve fitting methods.  In all three
tables we give the differences of the derived distance moduli and the
corresponding values in an empty universe ($\Omega_M = 0.0, \Omega_{\Lambda} =
0.0$).

Since the light curve decline of SN~2003jo was covered by the ground based
data, we also fit this object without the $HST$ data.  Using just the CTIO $RI$
data, the method of \citet{Pri_etal05} gives \dmm\ = 0.84 $\pm$ 0.17, E($B-V$)
= 0.05 $\pm$ 0.05, and ($m-M$) = 42.77 $\pm$ 0.20 (on an H$_0$ = 65 scale).  
These values are consistent with the solution given in Table \ref{lc_fits},
which used the combined CTIO 4-m and $HST$ data.  

The \dmm\ values given in Table \ref{lc_fits} indicate that all but one of
the objects discussed here are slow decliners.  The slowest declining
template object in the Prieto et al. training set is SN~1999aa,
with \dmm\ = 0.81 $\pm$ 0.04.  Thus, our objects are near, but not beyond,
the limit of the \dmm\ system.

The MLCS fits were done in two ways. First, we assumed a prior that no light
curve could be slower than MLCS $\Delta=-0.4$, the mimimum value of the MLCS
training set.  In this case we found that the data systematically deviated
from the fits at late times.  The actual light curves were slower than the
slowest declining objects in the training set.  We then fit the light curves
with no constraint on the possible values of $\Delta$.  In this case we had
to extrapolate beyond the training set. In \S3.4 we discuss the
cosmological effects of using the prior or not using it.

In Figure \ref{fits} we show the \dmm\ light-curve fits in the 
rest-frame bands for 7 SNe.  In Figure \ref{f011_iband} we also show
the K-corrected, extinction-corrected $I$-band data of SN~2003lh,
our only data obtained with NICMOS.  For comparison we show the
$I$-band data of SN~1999aa \citep{Kri_etal00,Jha02}, the slowest decliner 
in the nearby sample used by \citet{Pri_etal05}, and the $I$-band data of 
the prototypical slow decliner SN~1991T \citep{Lir_etal98}. 
The photometry of SN~1999aa has been adjusted 
in magnitude space to the brightness of SN~2003lh using the \dmm\ distance
modulus given in Table \ref{lc_fits} and 
assuming an absolute magnitude $M_I$(max) = $-$19.1 for the slowest-declining
SN~Ia studied by \citet{Nob_etal05}.  For $H_0$ = 65 \kms\ Mpc$^{-1}$, the
value used for the \dmm\ system, this becomes $M_I$ = $-$19.32 mag.  
The maximum of SN~1991T is made to coincide with that of SN~1999aa.  At face
value the $I$-band secondary hump of SN~2003lh was weaker than that of
SNe~1991T and 1999aa.  As we pointed out in a previous paper 
\citep[][Fig.18]{Kri_etal01}, SNe~Ia with identical decline rates in
$B$ and $V$ can have significantly different $I$-band secondary maxima.
Whatever is the appropriate adjustment of the SN~2003lh photometry in Figure
\ref{f011_iband}, we can say that its secondary $I$-band hump
occurred earlier than that of SN~1999aa.

As a measure of our systematics, we show in Figure \ref{mlcs_resids} the
residuals of the K-corrected CTIO 4-m and $HST$ data compared to the
rest-frame MLCS2k2 light-curve fits.  In Figure \ref{dm15_resids} we show
analogous plots of the residuals of the \dmm\ fits shown in Figure
\ref{fits}.  In these two figures 
a differential magnitude greater than zero means that
a rest-frame datum is fainter than the fit, and a differential magnitude less
than zero means that a rest-frame datum is brighter than the fit.
  
Figures \ref{mlcs_resids} and \ref{dm15_resids} show that there are no
statistically significant trends in the residuals of the K-corrected data
compared to the light-curve fits, and that the mean residual is close to
zero, as expected.

For the individual spectra of the SNe~Ia discussed in this paper, see
\cite{Mat_etal05}.  In an attempt to find spectral peculiarities of the slow
decliners consistent with their light curves, we created a composite spectrum
of SNe 2003jo, 2003kp, 2003kv, 2003lh, 2003le, and 2003li (see Figure
\ref{composite}).  We first deredshifted
each spectrum to its rest frame, then averaged each wavelength bin, using
only the spectra which covered that particular wavelength bin.  Using the
dates of $B$-band maximum in Table \ref{mlcs_fits}, the composite spectrum
has a mean spectral age of +5 days with respect to maximum light.

The composite spectrum shows features typical of a normal SN~Ia slightly past
T($B_{max}$).  The \ion{Ca}{2} H\&K lines are present and as strong as in the
normal SN~Ia 1992A at a similar age, unlike the over-luminous, slowly
declining SN~1991T \citep{Fil_etal92}.  The spectrum of the slowly declining
SN~1999aa evolved similarly to SN~1991T, with the major exception of having
strong \ion{Ca}{2} H\&K.  At a similar age to the composite spectrum,
SN~1999aa does not appear to be drastically different than the composite
spectrum.

Despite the increase of S/N in the ESSENCE objects by combining them into one
composite spectrum, the overall S/N is lower than the S/N obtained for the
spectra of many low-$z$ SNe.  We note, however, that the $\lambda$4130 feature
due to \ion{Si}{2} is weak and more like SN~1999aa than SN~1992A, consistent
with the slowly declining light curve fits found for these SNe. In a future
paper we shall provide a full length discussion of composite ESSENCE spectra
\citep{Fol_etal06}.

\subsection {Decline Rates of Supernovae}

In Figure \ref{dm15_hist} we show a histogram of \dmm\ values for 107 nearby
SNe~Ia, dividing them into two groups according to the host-galaxy type
\citep{Gal_etal05}.  Of these 107 hosts, 71 are spirals and 36 are
ellipticals or S0 galaxies.  We also add to the histogram seven values for
ESSENCE SNe listed in Table \ref{lc_fits}.  Figure \ref{dm15_hist} shows that
spiral galaxies are more likely to produce slowly declining SNe~Ia, while
ellipticals are more likely to produce rapidly declining ones; for further
details see \citet{Ham_etal96a} and \citet{Gal_etal05}.  \citet{Ume_etal99}
discuss why this might be the case, but for now it must just stand as an
empirical fact.  The striking aspect of the histogram is that all but one
of the ESSENCE SNe are at the extreme slow-declining end of the \dmm\
distribution.

Figure \ref{stamps} suggests that SN~2003jo, SN~2003ll, and SN~2003li
occurred in spiral galaxies on the basis of their morphology in the $HST$
images. SN~2003kv is projected near a bright galaxy, but spectroscopy is
needed to show that it is the host.  The other five SNe discussed here
occurred in very faint hosts, about which we have almost no information
for the purposes of morphological or spectroscopic classification.
None of the 9 SNe was found in a galaxy that is obviously an elliptical. 

Using the local sample as a guide, we would expect a small number (3)
of our ESSENCE sample to be in E/S0 galaxies.  But such an extrapolation
to a magnitude limited sample is dangerous.  SNe~Ia in nearby early-type
galaxies are, on average, fast decliners and intrinsically faint at 
maximum, so less likely to be found in our search.  Therefore, a lack of
early-type hosts is not surprising.  A more detailed analysis of this
bias is discussed below.

For a discussion of the properties of nearby galaxies that have hosted SNe
Ia, see \citet{Gal_etal05}. The morphology of the host galaxies of 
high-$z$ SNe~Ia has been discussed by a number of authors.  \citet{Far_etal02}
studied 22 galaxies at $z \approx 0.6$ and found that $\sim$70\% of
high-$z$ SNe occur in spirals and $\sim$30\% occur in 
ellipticals,\footnote[22]{At least one galaxy in the Farrah et al. sample
(the host of SN~1998M) is mis-classified as an elliptical.  Our multi-band
imagery indicates that the host is a blue star-forming galaxy.}
similar to the percentages for the local sample. They found no evidence
that SNe~Ia are preferentially found in the outer regions of the hosts, implying that
host galaxy extinctions of the high-$z$ sample should be comparable to the
local examples.  \citet{Wil_etal03} found no correlations of the distance
residuals with host-galaxy properties in the redshift range $0.42 < z <
1.06$. Their 18 galaxies which hosted SNe were discovered by the High-$z$
Supernova Search Team \citep{Schmidt_etal98, Rie_etal98}.  
\citet{Sul_etal03} found from a larger sample \citep[42 objects discovered
by the Supernova Cosmology Project]{Per_etal99} that there is evidence for
a larger {\em scatter} of the absolute magnitudes for the high-$z$ SNe
occurring in spirals.  They also found that SNe occurring in spirals are
on average marginally less luminous than those in E/S0 galaxies, by $0.14
\pm 0.09$ mag.  But this is opposite of what is seen at low redshifts
\citep{Ham_etal96a}.

These analyses do not lead us to believe that there are significant
differences between the hosts of low-$z$ and our high-$z$ SNe~Ia or
between the SNe~Ia themselves.  Why, then, do all but one of the ESSENCE
objects discussed in this paper have such slow decline rates?  Some of
the issues involved have already been discussed by \citet{Li_etal01}.  
It is important to note that the SNe~Ia in this paper not typical of
ESSENCE SNe in general.  They were chosen to be at the highest redshift
end of the ESSENCE distribution for $HST$ follow-up.

In Figure \ref{control_time} we provide an illustrative answer.  Let us 
assume that we discover all the SNe in the $R$ band, and ignore the $I$
band. Assume a fixed magnitude limit for detection; we shall use $R$ =
23.0 mag, which gives S/N $\approx 10$ for images obtained in good seeing
($0.9''$).  Varying the limit will change the details but not the general
result.  We calculate the time a SN~Ia stays above the detection limit as
a function of \dmm\ and redshift.  This is the ``control time,'' but we
will only include the time before maximum, since detection before the
time of maximum was a requirement for $HST$ follow-up observations.  
Figure \ref{control_time} shows that the control time is a steeply
falling function of \dmm . For $z < 0.5$ this is not a problem for the
ESSENCE search since we visit the same piece of sky every four nights.
But beyond $z \approx 0.6$ the fast-declining events are not above our
detection threshold long enough to have been included in the $HST$
sample. As we push to high redshift it is clear that our $HST$ selection
is highly biased to the slowest-declining SNe~Ia.

The selection effect which increased our sample of slowly declining light
curves relative to the average distribution will affect their cosmological
use.  Our SNe~Ia may be slower than average for their luminosity or
intrinsically brighter than expected for their observed light curve shape as
a result of the selection effect.  While this selection bias is strong in our
$HST$ sample, it will be present in the highest-redshift slice of any
magnitude-limited search for SNe~Ia. Fast-declining events are not only
fainter than slowly declining SNe~Ia, but are above any threshold for less
time, leading to a control-time bias against their discovery. In designing a
search that minimizes systematic errors, care must be taken to avoid this
bias by either visiting fields at a rapid cadence or pruning the
highest-redshift end of the accumulated sample.  Such a selection effect
should be modeled once we have our full sample of 200 SNe~Ia to fully remove
its impact.

\subsection{The Strange Case of SN~2003ku}

The spectral cross-correlation program SNID indicated that the spectrum of
SN~2003ku was most consistent with that of a SN~Ia with $z = 0.79$, but
redshift 0.41 was almost as likely \citep{Mat_etal05}. In Figure
\ref{e315_spec} we show these two possibilities.  SN 2003ku was the brightest
of the nine SNe discussed in this paper (see Figs. \ref{R_light_curves} and
\ref{I_light_curves}); thus, photometric considerations alone would favor the
smaller redshift. Using the method of \citet{Pri_etal05}, our attempts to
K-correct and fit the photometry of SN~2003ku to rest-frame $B$ and $V$
magnitudes gave $\chi_{\nu}^2$ = 2.8 for the lower redshift and $\chi_{\nu}^2$
= 5.2 for the larger value, implying that the lower redshift is favored, but
the fit is still not very good.  If $z$ = 0.79, this object gives a derived
distance modulus 2.5 mag ``too close'' compared to the empty-universe model.
The BATM analysis gave similar results; if $z$ = 0.79, the derived distance
modulus is 1.7 mag smaller than one would get in an empty-universe model.

While SNe exhibit some dispersion in luminosities, and light-curve fitting of
low S/N photometry adds to the scatter of measurable parameters, something is
clearly amiss with an object that is discrepant by 2 mag.  For SN~2003ku
several possibilities can be considered:

(1) The photometry contains some serious calibration error. One of the fields
represented in Figure \ref{RI_test} was the SN~2003ku field.  There is a
reasonable match, certainly better than 2 mag, of the $RI$ photometry and the
$HST$ photometry. We believe the photometry of this object is correct.  We
note that SN~2003ku was the only object discussed here whose $R$ and F625W
photometry {\em clearly} differ, by $\sim$0.4 mag. See Figure
\ref{R_light_curves}. From a cursory inspection of Figures \ref{e315_spec},
\ref{hst_filters}, and \ref{CTIO4m_filters} one can see that the local peak in
the spectrum at 7100--7300~\AA\ would be included in $R$-band photometry but
excluded from F625W photometry.  

(2) Could SN~2003ku be a gravitationally lensed object?  \citet{Hol01}
indicates that 0.05\% of sources at $z = 1$ are expected to be multiply imaged
on arcsecond scales.  \citet{Por_Mad00} investigated the lensing magnification
of high-redshift SNe for a variety of scenarios and found that up to 10\%
could be magnified by 0.1 to 0.3 mag at $z = 1$.  Further considerations are
elaborated by \citet[and references therein]{Wan05}.  At a redshift of 0.4 to
0.8 we conclude that the probability of a 2 mag magnification is extremely
small.  Also, there is no evidence of multiple images.

(3) Was SN~2003ku a spectroscopically peculiar SN~Ia?  This would also help
explain the confusion on the part of SNID to determine the redshift. 

(4) If it were a SN~Ia at $z \approx 0.4$ or less, then it would not have been
so overluminous.

(5) It could have been a SN~Ia with a different explosion mechanism.
\citet{Wil_Mat04}, for example, describe scenarios whereby white-dwarf stars
passing close to black holes of a range of masses (10 to 10$^9$ M$_{\odot}$)
explode.\footnote[23]{\citet{Kho_etal93} and \citet{Die_etal97} previously
described the interactions and disruption of stars, in particular n = 1.5
polytropes, with black holes.} These authors estimate that between 10$^{-2}$
and 10$^{-4}$ of the SNe~Ia out to $z = 1$ could be white dwarfs disrupted by
black holes.  Most significantly, they suggest that the light curves would
have different shapes.  We wonder if white dwarfs disrupted by black holes
might also have different spectra, or luminosities which differ from the
standard white-dwarf-plus-donor-star scenario.

(6) It could have been an extremely luminous SN~Ib or SN~Ic like SN~1992ar
\citep{Clo_etal00}, and with a spectrum different from any in SNID's database
of spectra.  We feel this is the mostly likely possibility.

In any case, given the ambiguity of the redshift (hence luminosity) and
uncertainty in its classification, SN~2003ku will be eliminated from our final
determination of the equation-of-state parameter.

\subsection{Cosmological Consequences}

We consider three cosmological models: the empty universe
($\Omega_M = 0.0, \Omega_{\Lambda} = 0.0$), the open universe
($\Omega_M = 0.3, \Omega_{\Lambda} = 0.0$), and the concordance model
($\Omega_M = 0.3, \Omega_{\Lambda} = 0.7$). In Figure \ref{diff_hub} we
show a differential Hubble diagram derived from the light-curve fits,
and compare the results to these three models.  For each SN shown
in the plot we determined the distance modulus using MLCS2k2
and subtracted off the distance modulus one would
get in an empty universe.  We used the 157 SNe in the ``gold''
set of \citet{Rie_etal04} and also included seven ESSENCE objects
discussed here.
Because the redshifts of SN~2003ku and e510 are uncertain or unknown, we do
not include them in the plot.  Consider Figure \ref{diff_hub}.  In
an empty universe the data points would equally likely fall above
and below the horizontal line in the middle, but it is obvious
that the majority of the points at $z \approx 0.5$ and $z \approx 0.8$
are above this. The dashed line corresponds to the
concordance model, while the dotted line corresponds to the open
universe with $\Omega_M = 0.3$. There is a half-magnitude range
(a root-mean-square uncertainty of $\pm$ 0.18 mag)
in the nearby sample ($z < 0.10$).  The more distant objects do not
show a {\em considerably} greater range, which is reassuring, since
we are assuming that SNe~Ia with lookback times of several
billion years are similar to those observed nearby.

In Figure \ref{diff_hub} the weighted mean difference of the ESSENCE
data points, compared to the open-universe model (i.e., dotted
line), is $\langle \Delta(m-M) \rangle = +0.37 \pm 0.09$ mag.
At $z = 0.6$ the difference between the concordance model (dashed 
line) and the open-universe model is +0.23 mag.  In a flat universe,
as $\Omega_M$ becomes smaller and $\Omega_{\Lambda}$ increases, the
curve bows upward at $z$ = 0.6.  Thus, if the
geometry of the Universe is flat, the ESSENCE data alone
would stipulate that the mass density of the Universe is 
$\Omega_M < 0.3$ and the dark energy has $\Omega_{\Lambda} > 0.7$.  

Using the MLCS2k2 fits and a nearby sample to establish the zeropoint of the
distance moduli, we obtain cosmological constraints for a cosmological
constant based on the ESSENCE supernovae.  In Figure
\ref{contour_local_essence} we have plotted constraints assuming a prior
limiting the decline rate of the slowest supernovae, and constraints based on
no prior which allow an extrapolation by MLCS. Clearly the choice of a prior
affects the cosmological results and points to the need to avoid a selection
bias that preferentially discovers slowly declining supernovae at the limits
of the decline rate range of the local sample.  Using the 2dF mass constraint
of \citet{Col_etal05} ($\Omega_M h$ = 0.168 $\pm$ 0.016), the Hubble constant
of \citet{Fre_etal01}, and with the prior on MLCS $\Delta$ we find
$\Omega_{\Lambda}$ = 1.26 $\pm$ 0.18 for the sample of ESSENCE plus nearby
SNe.  Without the prior we find $\Omega_{\Lambda}$ = 0.99 $\pm$ 0.21.

In Figures \ref{contour_lambda} and \ref{contour_w} we show constraints
obtained for $\Omega_{\Lambda}$ and $w$ using
the entire ``gold'' set of 157 SNe from \citet{Rie_etal04} plus seven ESSENCE
SNe discussed in this paper. We have assumed $w$ = $-$1 for Figure
\ref{contour_lambda} and have used two different matter constraints: the 2dF
mass constraint of \citet{Col_etal05} coupled with
the Hubble constant from \citet{Fre_etal01}, and the SDSS large-scale
structure result of $\Omega_M$ = 0.273 $\pm$ 0.025 + 0.137$\Omega_K$ 
\citep{Eis_etal05}.  Figure \ref{contour_lambda} shows that the standard
model is recovered
at the 1$\sigma$ level if we use the full ``gold'' SN sample, the
seven ESSENCE objects from Table \ref{mlcs_fits}, and either matter constraint.
From Figure \ref{contour_w} we find $w \; = \; -0.88 \pm 0.11$ with the
prior on MLCS $\Delta$ for the ESSENCE sample.  Without the prior we find
$w \; = \; -0.90 \pm 0.12$.  This is a significant shift in $w$ considering 
only 7 of 164 SNe~Ia were affected by the change of prior.  This demonstrates
the importance of the selection bias at the high-$z$ end of any survey.


\section{Conclusions}

We have presented photometry of nine supernovae from the ESSENCE
project.  Ground-based photometry allowed us to cover the maxima in
the rest-frame $B$ and $V$ light curves of all the objects discussed
here, while the photometry obtained with $HST$/ACS allowed us to
characterize the light-curve tails. The light-curve fitting of seven
objects with reliable redshifts, carried out with three different
methods, gave distance moduli consistent within the errors.

On the basis of the values of \dmm\ derived for the ESSENCE SNe, all but
one are slow decliners compared to the local sample and their
light curves are as slow as the slowest found in the local set
of well-observed supernovae. We show that the SNe selected
to be observed with $HST$ were at the high-$z$ end of our distribution of
redshifts and probably represent a selection bias for slow-declining
events. 

The ESSENCE project is being carried out every other night on the
CTIO 4-m telescope during the months of October, November, and
December.  After three years of the ESSENCE project we have
discovered roughly 100 SNe and SN candidates. Eventually, with
$\sim$200 spectroscopically confirmed ESSENCE SNe~Ia, all observed
with the same ground-based telescope and filter system, we should be
able to determine the time average of the equation-of-state
parameter of the Universe 
to $\pm$10\%.

We observe two sets of fields on alternating observing nights.  The
resulting cadence of light-curve points every four observer-frame days is
clearly sufficient to characterize the light curves. For SNe~Ia with $z
\approx 0.5$, we obtain apparent magnitudes at maximum with an uncertainty
of $\pm$0.06 mag.  For further details on the project strategy see
\citet{Mik_etal05}.

The 157 ``gold'' SNe~Ia of \citet{Rie_etal04}, along with seven
high-redshift SNe discussed here, give contours in the $\Omega_M -
\Omega_{\Lambda}$ plane consistent with a positive cosmological constant
and flat geometry. Further cosmological tests await the acquisition of
larger self-consistent data sets.

\acknowledgments

The ESSENCE Project is supported primarily by NSF grants AST-0206329 and
AST-0443378. We are also grateful for NASA grants GO-9860 and AR-9925 from the
Space Telescope Science Institute (STScI), which is operated by AURA, Inc.,
under NASA contract NAS 5-26555.  P.M.G. is supported in part by NASA Long Term
Space Astrophysics grant NAGS-9364.  We thank Galina Soutchkova of STScI for her
help in scheduling our pseudo-TOO observations with $HST$.  Some of the results
presented herein were obtained at the W. M. Keck Observatory, which is operated
as a scientific partnership among the California Institute of Technology, the
University of California, and NASA; the Observatory was made possible by the
generous financial support of the W. M. Keck Foundation.  VLT observations were
part of program 170.A-0519.  We also utilized the SDSS early data release.
A.V.F. is grateful for the support of NSF grant AST-0307894, and for a Miller
Research Professorship at UC Berkeley during which part of this work was
completed. We thank Jorge Araya for tracing filters in the lab, Sean Points for
further analysis of those traces, and George Jacoby for providing his program
for simulating the filter-transmission profiles of the $RI$ filters appropriate
to the focal ratio of the CTIO 4-m telescope. We acknowledge Joseph Gallagher
for providing his database of supernova properties. Lou Strolger kindly derived
the F110W photometry of one of our supernovae.  K.K. thanks David Spergel for
many stimulating discussions.  Finally, we thank an anonymous referee for
constructive suggestions and references to other work.

\section{Appendix: Filters for Supernova Photometry}

In Figure \ref{hst_filters} we show the effective throughput (i.e., combination of 
filter transmission and quantum efficiency as a function of wavelength) of the
three filters used with $HST$/ACS.


We used laboratory hardware and software produced by Ocean Optics to make
trans\-mis\-sion-curve traces of the $R$-band and $I$-band filters, which
were used with the CTIO 4-m telescope and its facility mosaic camera.  
The traces were taken with the incidence angle of the input laser beam
ranging from $0^\circ$ to $11^\circ$. We subsequently used the data files
and a program kindly provided by G. Jacoby to simulate the filter traces
appropriate for the $f/2.7$ beam of the CTIO 4-m telescope and its Mosaic
camera.  As the $I$-band filter is an interference filter, this is
particularly important. The effective $I$-band filter profile for an
$f/2.7$ beam has half-power points shifted roughly 30 \AA\ toward the blue
from the $0^\circ$ incidence angle tracing.

The $RI$ filter transmission curves can be obtained in
graphical and tabular form at website http://www.ctio.noao.edu/$\sim$points/FILTERS.
Our effective $RI$ filter transmission curves are shown in Figure \ref{CTIO4m_filters}.
They include the effects of reflection off the primary mirror, the quantum 
efficiency of the CCD chips as a function of wavelength, the atmospheric extinction,
and the major telluric absorption lines.


\begin{deluxetable}{lcllcc}
\tablewidth{0pt}
\tablecolumns{6}
\tablecaption{High-Redshift Supernovae: Basic Data\label{obstab}}
\tablehead{
\colhead{IAU name} &
\colhead{ESSENCE} &
\colhead{$\alpha$(J2000)} &
\colhead{$\delta$(J2000)} &
\colhead{Redshift$^a$} & \colhead{$E(B-V)_{Gal}^b$}
}
\startdata
SN~2003jo & d033 & 23:25:24.03 & $-$09:26:00.6 & 0.53 & 0.036  \\

SN~2003kp & e147 & 02:31:02.64 & $-$08:39:50.8 & 0.64 & 0.032   \\
SN~2003ku & e315 & 01:08:36.25 & $-$00:33:20.8 & 0.79\tablenotemark{c} & 0.036   \\
\nodata   & e510 & 23:30:59.97 & $-$08:37:34.4 & 0.68\tablenotemark{d} & 0.032 \\
SN~2003kv & e531 & 02:09:42.52 & $-$03:46:48.6 & 0.78 & 0.023   \\

SN~2003lh & f011 & 02:10:19.51 & $-$04:59:32.3 & 0.54 & 0.020   \\
SN~2003le & f041 & 01:08:08.73 & +00:27:09.7   & 0.56 & 0.029   \\
SN~2003ll & f216 & 02:35:41.19 & $-$08:06:29.6 & 0.60 & 0.033  \\
SN~2003li & f244 & 02:27:47.29 & $-$07:33:46.2 & 0.54  & 0.027  \\
\enddata
\tablenotetext{a} {Obtained from the spectra of the SNe themselves, rather
than the host galaxies.  We used our program SNID (Tonry et al. 2005, in 
preparation).}
\tablenotetext{b} {From the reddening maps of \citet{Sch_etal98}; magnitude units.}
\tablenotetext{c} {See text and \citet{Mat_etal05} for a discussion of the
redshift of this object.}
\tablenotetext{d} {The minimum reduced $\chi^2$ value of the light-curve 
fits is obtained for $z = 0.68$; no spectroscopic redshift was obtained.}
\end{deluxetable}


\begin{deluxetable}{ccllll}
\tablewidth{0pt}
\tablecolumns{6}
\tablecaption{$HST$ Photometry\label{hstphot}\tablenotemark{a}}
\tablehead{
\colhead{SN name} &
\colhead{JD\tablenotemark{b}} &
\colhead{F625W} &
\colhead{F775W} &
\colhead{F850LP} &
\colhead{F110W}
}
\startdata
2003jo	& 2946.95 & 23.031 (0.018)	& 22.661 (0.017) & 22.455 (0.023) & \nodata \\
\ldots  & 2953.01 & 23.399 (0.022)	& 22.841 (0.018) & 22.748 (0.035) & \nodata \\
\ldots  & 2960.27 & \nodata     	& 23.201 (0.019) & \nodata        & \nodata \\
\ldots  & 2973.81 & \nodata      	& 23.799 (0.028) & 23.186 (0.100) & \nodata \\
\ldots  & 2976.41 & 25.045 (0.065)	& 23.863 (0.035) & 23.210 (0.035) & \nodata \\
\ldots  & 2985.99 & \nodata     	& 24.277 (0.040) & 23.464 (0.036) & \nodata \\
\ldots  & 3148.26 & 26.75 (+0.51/$-$0.35) & 26.91 (+0.35/$-$0.26) & \nodata & \nodata \\
\hline

2003kp	& 2981.89 & 23.540 (0.024) & 23.006 (0.020) & 22.789 (0.027) & \nodata \\
\ldots  & 2988.67 & 23.961 (0.034) & 23.394 (0.025) & 23.081 (0.032) & \nodata \\
\ldots	& 2995.16 & \nodata	   & 23.657 (0.024) & 23.282 (0.042) & \nodata \\
\ldots	& 3007.67 & \nodata	   & 24.198 (0.037) & 23.576 (0.038) & \nodata \\
\ldots	& 3021.02 & \nodata	   & 24.766 (0.053) & 23.913 (0.049) & \nodata \\
\hline

2003ku  & 2976.02 & 23.145 (0.027) & 22.249 (0.015) & 21.904 (0.018) & \nodata \\
\ldots  & 2983.15 & 23.614 (0.026) & 22.550 (0.017) & 22.009 (0.018) & \nodata \\
\ldots	& 2989.67 & \nodata	   & 22.945 (0.023) & 22.114 (0.017) & \nodata \\
\ldots	& 3001.88 & \nodata	   & 23.601 (0.025) & 22.522 (0.021) & \nodata \\
\ldots	& 3016.29 & \nodata	   & 24.168 (0.053) & 23.200 (0.032) & \nodata \\
\hline

e510	& 2983.08 & 23.799 (0.043) & 23.413 (0.027) & 23.315 (0.038) & \nodata \\
\ldots	& 2989.75 & \nodata	   & 23.723 (0.028) & 23.524 (0.026) & \nodata \\
\ldots	& 3001.53 & \nodata	   & 24.474 (0.051) & 23.824 (0.051) & \nodata \\
\ldots	& 3016.15 & \nodata	   & 25.168 (0.095) & 24.298 (0.110) & \nodata \\
\hline

2003kv	& 2981.82 & 24.277 (0.042) & 23.228 (0.024) & 23.149 (0.037) & \nodata \\
\ldots  & 2988.81 & 24.813 (0.065) & 23.518 (0.030) & 23.379 (0.044) & \nodata \\
\ldots	& 2995.09 & \nodata	   & 23.896 (0.031) & 23.559 (0.040) & \nodata \\
\ldots	& 3006.82 & \nodata	   & 24.510 (0.049) & 24.012 (0.059) & \nodata \\
\ldots	& 3020.95 & \nodata	   & 25.265 (0.093) & 24.592 (0.094) & \nodata \\
\hline

2003lh	& 3009.76 & 23.877 (0.031) & 23.060 (0.021) & 22.979 (0.030) & \nodata \\
\ldots	& 3015.96 & 24.325 (0.060) & 23.363 (0.025) & 23.119 (0.046) & \nodata \\
\ldots  & 3016.60 & \nodata        &   \nodata      &   \nodata      & 23.292 (0.026) \\
\ldots	& 3022.89 & \nodata	   & 23.704 (0.026) & 23.179 (0.030) & \nodata \\
\ldots  & 3024.36 & \nodata        &   \nodata      &   \nodata      & 23.400 (0.028) \\
\ldots  & 3030.60 & \nodata        &   \nodata      &   \nodata      & 23.463 (0.021) \\
\ldots	& 3035.89 & \nodata	   & 24.208 (0.037) & 23.624 (0.040) & \nodata \\
\ldots	& 3042.75 & \nodata	   & 24.430 (0.044) & 23.888 (0.067) & \nodata \\
\hline

2003le	& 3003.01 & 22.899 (0.017) & 22.529 (0.017) & 22.407 (0.022) & \nodata \\
\ldots	& 3009.68 & 23.299 (0.022) & 22.725 (0.018) & 22.603 (0.025) & \nodata \\
\ldots	& 3016.68 & \nodata	   & 22.983 (0.018) & 22.814 (0.025) & \nodata \\
\ldots	& 3029.68 & \nodata	   & 23.501 (0.025) & 23.048 (0.030) & \nodata \\
\ldots	& 3042.81 & \nodata	   & 24.084 (0.041) & 23.372 (0.055) & \nodata \\
\hline

2003ll  & 3004.71 & 24.593 (0.051) & 23.827 (0.033) & 23.174 (0.033) & \nodata \\
\ldots  & 3011.65 & 25.351 (0.139) & 24.529 (0.055) & 23.452 (0.042) & \nodata \\
\ldots  & 3018.44 & \nodata        & 24.570 (0.065) & 23.496 (0.035) & \nodata \\
\ldots  & 3029.75 & \nodata        & 25.070 (0.135) & 23.761 (0.040) & \nodata \\
\ldots  & 3043.71 & \nodata        & 25.336 (0.083) & 24.356 (0.156) & \nodata \\
\hline

2003li	& 3011.22 & 23.983 (0.047) & 23.121 (0.022) & 23.094 (0.034) & \nodata \\
\ldots  & 3018.82 & 24.544 (0.050) & 23.459 (0.027) & 23.172 (0.035) & \nodata \\
\ldots	& 3025.95 & \nodata	   & 23.735 (0.037) & 23.334 (0.046) & \nodata \\
\ldots	& 3036.20 & \nodata	   & 24.244 (0.038) & 23.577 (0.040) & \nodata \\
\ldots	& 3050.60 & \nodata	   & 24.705 (0.057) & 23.956 (0.053) & \nodata \\
\enddata
\tablenotetext{a} {The ACS magnitudes given are ``Vega'' magnitudes derived
using 50-pixel-radius zeropoints \citep{Sir_etal05}, which are based on
the Vega spectrophotometric calibration of \citet{Boh_Gil04}. The values
in parentheses are 1$\sigma$ error bars.  The F110W photometry was obtained
with NICMOS.}
\tablenotetext{b} {Julian Date minus 2,450,000.}
\end{deluxetable}

\begin{deluxetable}{ccll}
\tablewidth{0pt}
\tablecolumns{5}
\tablecaption{Ground-Based $R$ and $I$ Photometry\label{ctio4mphot}\tablenotemark{a}}
\tablehead{
\colhead{SN name} &
\colhead{JD\tablenotemark{b}} &
\colhead{$R_{nat}$} &
\colhead{$I_{nat}$}
}
\startdata
2003jo & 2931.52 &  22.63 ($\pm$0.06)     & 22.57 (+0.13/$-$0.12) \\
\ldots & 2934.51 &  22.63 ($\pm$0.10)     & 22.56 (+0.13/$-$0.12) \\
\ldots & 2940.54 &  22.76 ($\pm$0.08)     & 22.74 (+0.16/$-$0.14) \\
\ldots & 2944.51 &  23.03 (+0.20/$-$0.17) &  \nodata              \\
\ldots & 2958.53 &  23.77 (+0.27/$-$0.22) & 23.07 (+0.64/$-$0.40) \\
\ldots & 2962.52 &  24.14 (+0.52/$-$0.35) &  \nodata              \\
\ldots & 2966.54 &  24.07 (+0.57/$-$0.37) & 23.19 (+0.21/$-$0.18) \\
\ldots & 2970.59 &  24.46 (+0.41/$-$0.30) & 23.10 (+0.22/$-$0.19) \\
\ldots & 2972.56 &  24.30 (+0.28/$-$0.23) & 23.83 (+0.88/$-$0.48) \\
\ldots & 2976.56 &  24.49 (+0.90/$-$0.49) & 23.39 (+0.26/$-$0.21) \\
\ldots & 2986.55 &  $>$24.81              & 23.79 (+0.40/$-$0.29) \\
\ldots & 2990.54 &  \nodata               & 23.80 (+0.65/$-$0.40) \\
\ldots & 2994.55 &  \nodata               & 24.83 (+1.68/$-$0.63) \\
\hline

2003kp & 2936.65 &  $>$ 24.90             &   \nodata             \\
\ldots & 2942.66 &  24.09 (+0.26/$-$0.21) & 24.21 (+1.18/$-$0.55) \\
\ldots & 2944.62 &  23.84 (+0.35/$-$0.26) & 23.43 (+0.20/$-$0.17) \\
\ldots & 2960.68 &    \nodata             & 22.41 (+0.31/$-$0.24) \\
\ldots & 2964.68 &  22.67 ($\pm$0.06)     & 22.42 ($\pm$0.12)     \\
\ldots & 2968.67 &  23.20 (+0.56/$-$0.37) & 22.60 ($\pm$0.11)     \\
\ldots & 2970.66 &  22.96 ($\pm$0.08)     &  \nodata              \\
\ldots & 2970.68 &  22.92 ($\pm$0.05)     & 22.53 ($\pm$0.11)     \\
\ldots & 2972.67 &  23.07 ($\pm$0.06)     & 22.47 ($\pm$0.10)     \\
\ldots & 2974.62 &  23.14 ($\pm$0.10)     & 22.58 ($\pm$0.12)     \\
\ldots & 2976.63 &  23.21 ($\pm$0.09)     & 22.55 ($\pm$0.15)     \\
\ldots & 2988.68 &  24.24 (+0.24/$-$0.20) & 23.51 (+0.55/$-$0.36) \\
\ldots & 2992.68 &  23.80 (+0.24/$-$0.19) & 23.84 (+0.58/$-$0.38) \\
\ldots & 2996.67 &  24.30 (+0.47/$-$0.33) & 23.83 (+0.75/$-$0.44) \\
\ldots & 2998.65 &  24.41 (+0.73/$-$0.43) &  \nodata              \\
\hline

2003ku & 2934.57 &  $>$ 24.87             & $>$ 25.29             \\
\ldots & 2940.58 &  $>$ 24.93             & $>$ 24.94             \\
\ldots & 2958.58 &  23.06  ($\pm$0.14)    & 23.03 ($\pm$0.18)     \\
\ldots & 2962.59 &  22.94  ($\pm$0.09)    & 22.70 ($\pm$0.15)     \\
\ldots & 2972.59 &  22.57  ($\pm$0.07)    & 22.20 ($\pm$0.08)     \\
\ldots & 2974.55 &  22.58  ($\pm$0.04)    & 22.22 ($\pm$0.09)     \\
\ldots & 2986.59 &  23.26  ($\pm$0.07)    & 22.52 ($\pm$0.11)     \\
\ldots & 2990.57 &  23.55  ($\pm$0.16)    & 23.03 ($\pm$0.19)     \\
\ldots & 2994.58 &  23.94  ($\pm$0.14)    & 22.82 ($\pm$0.13)     \\
\ldots & 3000.59 &  24.15  ($\pm$0.17)    & 23.16 ($\pm$0.14)     \\
\hline

e510   & 2942.53 &      \nodata           & 24.05 (+0.32/$-$0.25) \\
\ldots & 2960.54 &      \nodata           & 23.14 (+0.38/$-$0.28) \\
\ldots & 2964.54 &  24.14 (+0.42/$-$0.30) & 23.06 (+0.28/$-$0.22) \\
\ldots & 2966.62 &  23.81 (+0.24/$-$0.20) & 22.95 (+0.25/$-$0.20) \\
\ldots & 2970.57 &  24.50 (+1.07/$-$0.53) & 23.12 (+0.21/$-$0.18) \\
\ldots & 2972.53 &  24.27 (+0.44/$-$0.31) & 23.04 (+0.17/$-$0.15) \\
\ldots & 2976.54 &  24.43 (+0.98/$-$0.51) & 23.39 (+0.50/$-$0.34) \\
\ldots & 2988.56 &      \nodata           & 23.70 (+0.52/$-$0.35) \\
\ldots & 2994.54 &  24.64 (+1.40/$-$0.59) & 23.89 (+0.48/$-$0.33) \\
\ldots & 2996.55 &  $>$ 25.36             & 23.90 (+0.38/$-$0.28) \\
\ldots & 2998.55 &      \nodata           & 24.02 (+1.24/$-$0.56) \\
\ldots & 3000.55 &      \nodata           & 24.34 (+0.85/$-$0.47) \\
\hline

2003kv & 2936.59 &  $>$ 23.80             &      \nodata          \\
\ldots & 2942.60 &  24.60 (+0.46/$-$0.32) & $>$  25.00            \\
\ldots & 2944.57 &  24.41 (+1.48/$-$0.60) & 23.37 (+0.44/$-$0.31) \\
\ldots & 2960.61 &  23.36 (+0.14/$-$0.12) &  \nodata              \\
\ldots & 2964.62 &  23.47 (+0.17/$-$0.15) & 22.86 (+0.19/$-$0.16) \\
\ldots & 2968.57 &  23.63 (+0.35/$-$0.27) & 23.35 (+0.24/$-$0.20) \\
\ldots & 2970.62 &  23.48 (+0.16/$-$0.14) & 23.08 (+0.29/$-$0.23) \\
\ldots & 2970.68 &  23.73 (+0.16/$-$0.14) &    \nodata            \\
\ldots & 2972.62 &  23.89 (+0.30/$-$0.23) & 22.83 (+0.17/$-$0.14) \\
\ldots & 2974.57 &  23.82 (+0.15/$-$0.13) & 23.11 (+0.20/$-$0.17) \\
\ldots & 2976.58 &  23.75 (+0.23/$-$0.19) & 23.26 (+0.27/$-$0.22) \\
\ldots & 2988.62 &  24.03 (+0.25/$-$0.20) & 23.94 (+1.03/$-$0.52) \\
\ldots & 2992.64 &  25.11 (+1.41/$-$0.59) &  \nodata              \\
\ldots & 2996.62 &  24.35 (+0.48/$-$0.33) & 24.36 (+0.90/$-$0.49) \\
\ldots & 2998.61 &  24.53 (+0.74/$-$0.43) & 23.88 (+0.41/$-$0.30) \\
\ldots & 3000.61 &  24.61 (+0.41/$-$0.30) & 24.51 (+0.87/$-$0.48) \\
\hline

2003lh & 2962.62 &  $>$ 25.54             & $>$ 24.57             \\
\ldots & 2968.62 &  23.82 (+0.26/$-$0.21) & 24.63 (+2.58/$-$0.70) \\
\ldots & 2986.62 &  22.53 ($\pm$0.07)     & 22.92 (+0.29/$-$0.23) \\
\ldots & 2990.60 &  22.71 ($\pm$0.06)     & 22.60 ($\pm$0.15)     \\
\ldots & 2994.61 &  22.75 ($\pm$0.07)     & 22.97 (+0.22/$-$0.18) \\
\ldots & 2996.64 &  23.00 ($\pm$0.10)     & 22.73 ($\pm$0.12)     \\
\ldots & 2998.63 &  22.90 ($\pm$0.08)     & 23.14 (+0.19/$-$0.16) \\
\ldots & 3000.62 &  23.17 ($\pm$0.08)     & 23.14 (+0.23/$-$0.19) \\
\hline

2003le & 2962.58 &  23.81 (+0.28/$-$0.23) &   \nodata             \\
\ldots & 2966.58 &  23.79 (+0.31/$-$0.24) & $>$  24.63            \\
\ldots & 2972.59 &  22.98 ($\pm$0.08)     & 23.05 ($\pm$0.12)     \\
\ldots & 2974.54 &  22.80 ($\pm$0.07)     & 22.72 ($\pm$0.10)     \\
\ldots & 2986.58 &  22.46 ($\pm$0.06)     & 22.29 ($\pm$0.08)     \\
\ldots & 2990.56 &  22.55 ($\pm$0.07)     & 22.17 ($\pm$0.07)     \\
\ldots & 2994.56 &  22.49 ($\pm$0.05)     & 22.20 ($\pm$0.11)     \\
\ldots & 2996.58 &  22.58 ($\pm$0.09)     & 22.44 ($\pm$0.11)     \\
\ldots & 2998.58 &  22.59 ($\pm$0.05)     & 22.39 ($\pm$0.13)     \\
\ldots & 3000.58 &  22.72 ($\pm$0.06)     & 22.42 ($\pm$0.08)     \\
\hline

2003ll & 2964.66 &  25.09 (+1.70/$-$0.63) & $>$ 24.90             \\
\ldots & 2968.65 &  24.42 (+1.23/$-$0.56) & \nodata               \\
\ldots & 2972.67 &  23.64 ($\pm$0.19)     & 24.09 (+0.59/$-$0.38) \\
\ldots & 2974.61 &  23.53 ($\pm$0.19)     & 22.47 ($\pm$0.08)     \\
\ldots & 2976.62 &  23.10 ($\pm$0.13)     & 22.58 ($\pm$0.13)     \\
\ldots & 2988.66 &  23.14 ($\pm$0.13)     & 23.60 (+0.81/$-$0.46) \\
\ldots & 2992.67 &  23.24 ($\pm$0.08)     & 24.05 (+0.75/$-$0.44) \\
\ldots & 2996.66 &  24.88 (+0.27/$-$0.21) &  \nodata              \\
\hline

2003li & 2958.65 &  $>$ 25.64             & $>$ 25.38             \\
\ldots & 2962.66 &  $>$ 25.57             & $>$ 24.87             \\
\ldots & 2966.66 &  24.76 (+0.34/$-$0.26) & 23.79 (+0.62/$-$0.39) \\
\ldots & 2986.66 &  22.75 ($\pm$0.06)     & 22.49 ($\pm$0.08)     \\
\ldots & 2990.64 &  22.69 ($\pm$0.07)     & 22.76 ($\pm$0.11)     \\
\ldots & 2994.64 &   \nodata              & 22.67 ($\pm$0.09)     \\
\ldots & 2996.66 &  23.08 ($\pm$0.11)     & 22.88 ($\pm$0.09)     \\
\ldots & 2998.65 &  23.15 ($\pm$0.09)     & 22.78 ($\pm$0.11)     \\
\enddata
\tablenotetext{a} {The values given are natural system ``Vega'' magnitudes.
See text for more details.
%
%
If a value is given as ``$>$ some number,'' it is a 1$\sigma$ upper limit.
The values in parentheses are 1$\sigma$ lower and upper error bars.}
\tablenotetext{b} {Julian Date minus 2,450,000.}
\end{deluxetable}

\begin{deluxetable}{lccccccc}
\tablewidth{0pt}
\tabletypesize{\small}
\tablecaption{MLCS Fits to Seven ESSENCE SNe~Ia\label{mlcs_fits}\tablenotemark{a}}
\tablehead{
\colhead{SN}  & 
\colhead{$z$}  & \colhead{$t(B_{max})$ \tablenotemark{b}} &
\colhead{MLCS $\Delta$} & \colhead{$A_V^c$ (mag)}  &
\colhead{$m-M$ (mag)} & \colhead{$\Delta$($m-M$)}
}
\startdata
2003jo &  0.53  & 2934.93 & $-$0.40(0.09) & 0.44(0.10) & 42.67(0.20) &   +0.20(0.20)  \\

2003kp &  0.64  & 2962.80 & $-$0.40(0.11) & 0.16(0.13) & 43.11(0.20) &   +0.14(0.20)  \\
2003kv &  0.78  & 2963.70 & $-$0.40(0.15) & 0.30(0.21) & 43.46(0.27) & $-$0.05(0.27)  \\

2003lh &  0.54  & 2984.76 & $-$0.40(0.14) & 0.06(0.14) & 43.19(0.20) &   +0.67(0.20)  \\
2003le &  0.56  & 2985.90 & $-$0.40(0.13) & 0.13(0.12) & 42.76(0.21) &   +0.15(0.21)  \\
2003ll &  0.60  & 2979.79 &   +0.52(0.28) & 0.28(0.26) & 42.08(0.52) & $-$0.72(0.52)  \\
2003li &  0.54  & 2986.11 & $-$0.40(0.14) & 0.24(0.14) & 43.03(0.24) &   +0.51(0.24)  \\
\enddata
\tablenotetext{a} {Using Version 3 of the Multi-color Light Curve Shape method, MLCS2k2
\citep{Jha_etal05}. These results allow no extrapolation beyond the training set.
The last column contains the
arithmetic differences of the derived distance moduli and those expected
in an empty universe ($\Omega_M = 0.0, \Omega_{\Lambda} = 0.0$).}
\tablenotetext{b} {Time of $B$-band maximum.  Julian Date minus 2,450,000.}
\tablenotetext{c} {Host-galaxy extinction.  The Galactic extinction has been
subtracted, using the color excesses given in Table \ref{obstab}.}
\end{deluxetable}

\begin{deluxetable}{lcccc}
\tablewidth{0pt}
\tabletypesize{\small}
\tablecaption{BATM Fits to Seven ESSENCE SNe~Ia\label{batm_fits}\tablenotemark{a}}
\tablehead{
\colhead{SN}  & \colhead{$z$}  &
\colhead{$A_V^b$ (mag)}  &
\colhead{$m-M$ (mag)} & \colhead{$\Delta$($m-M$)}
}
\startdata
2003jo &  0.53   & 0.21(0.25) & 42.89(0.31) &   +0.44(0.31)  \\

2003kp &  0.64   & 0.01(0.11) & 43.11(0.19) &   +0.16(0.19)  \\
2003kv &  0.78   & 0.02(0.12) & 44.12(0.21) &   +0.62(0.21)  \\

2003lh &  0.54   & 0.03(0.13) & 43.20(0.22) &   +0.70(0.22)  \\
2003le &  0.56   & 0.03(0.13) & 42.86(0.15) &   +0.26(0.15)  \\
2003ll &  0.60   & 0.60(0.32) & 43.03(0.53) &   +0.25(0.53)  \\
2003li &  0.54   & 0.17(0.26) & 43.03(0.35) &   +0.53(0.35)  \\
\enddata
\tablenotetext{a} {Using the Bayesian Adapted Template Method \citep[and Tonry et al. 2005, in preparation]
{Ton_etal03}.  The last column contains the
arithmetic differences of the derived distance moduli and those expected
in an empty universe ($\Omega_M = 0.0, \Omega_{\Lambda} = 0.0$).}
\tablenotetext{b} {Host-galaxy extinction.  The Galactic extinction has been
subtracted, using the color excesses given in Table \ref{obstab}.}
\end{deluxetable}

\begin{deluxetable}{lcccccc}
\tablewidth{0pt}
\tabletypesize{\small}
\tablecolumns{7}
\tablecaption{\dmm\ Fits to Seven ESSENCE SNe~Ia\label{lc_fits}\tablenotemark{a}}
\tablehead{
\colhead{SN} &
\colhead{$z$} &
\colhead{$\chi_{\nu}^2$} & 
\colhead{\dmm} & 
\colhead{$E(B-V)_{host}$} &
\colhead{$m-M$ (mag)} &
\colhead{$\Delta$($m-M$)}
}
\startdata
2003jo & 0.53  & 0.57 & 0.83(0.04) & 0.09(0.02) & 42.62(0.19) & +0.17(0.19)  \\
2003kp & 0.64  & 1.80 & 0.88(0.02) & 0.01(0.02) & 43.09(0.18) & +0.14(0.18)  \\


2003kv & 0.78  & 0.56 & 0.92(0.02) & 0.06(0.03) & 43.54(0.20) & +0.04(0.20)  \\

2003lh & 0.54  & 2.49 & 0.87(0.03) & 0.07(0.06) & 42.63(0.21) & +0.13(0.21)  \\
2003le & 0.56  & 1.56 & 0.84(0.04) & 0.06(0.02) & 42.51(0.19) & $-$0.09(0.19)  \\
2003ll & 0.60  & 3.35 & 1.30(0.02) & 0.08(0.07) & 42.62(0.29) & $-$0.16(0.29)  \\
2003li & 0.54  & 2.90 & 0.83(0.04) & 0.11(0.04) & 42.68(0.21) & +0.18(0.21)  \\
\enddata
\tablenotetext{a} {These values are from the light-curve fits using the 
\dmm\ method of \citet{Pri_etal05}.  The last column contains the
arithmetic differences of the derived distance moduli and those expected
in an empty universe ($\Omega_M = 0.0, \Omega_{\Lambda} = 0.0$).}
\end{deluxetable}

\clearpage

\figcaption[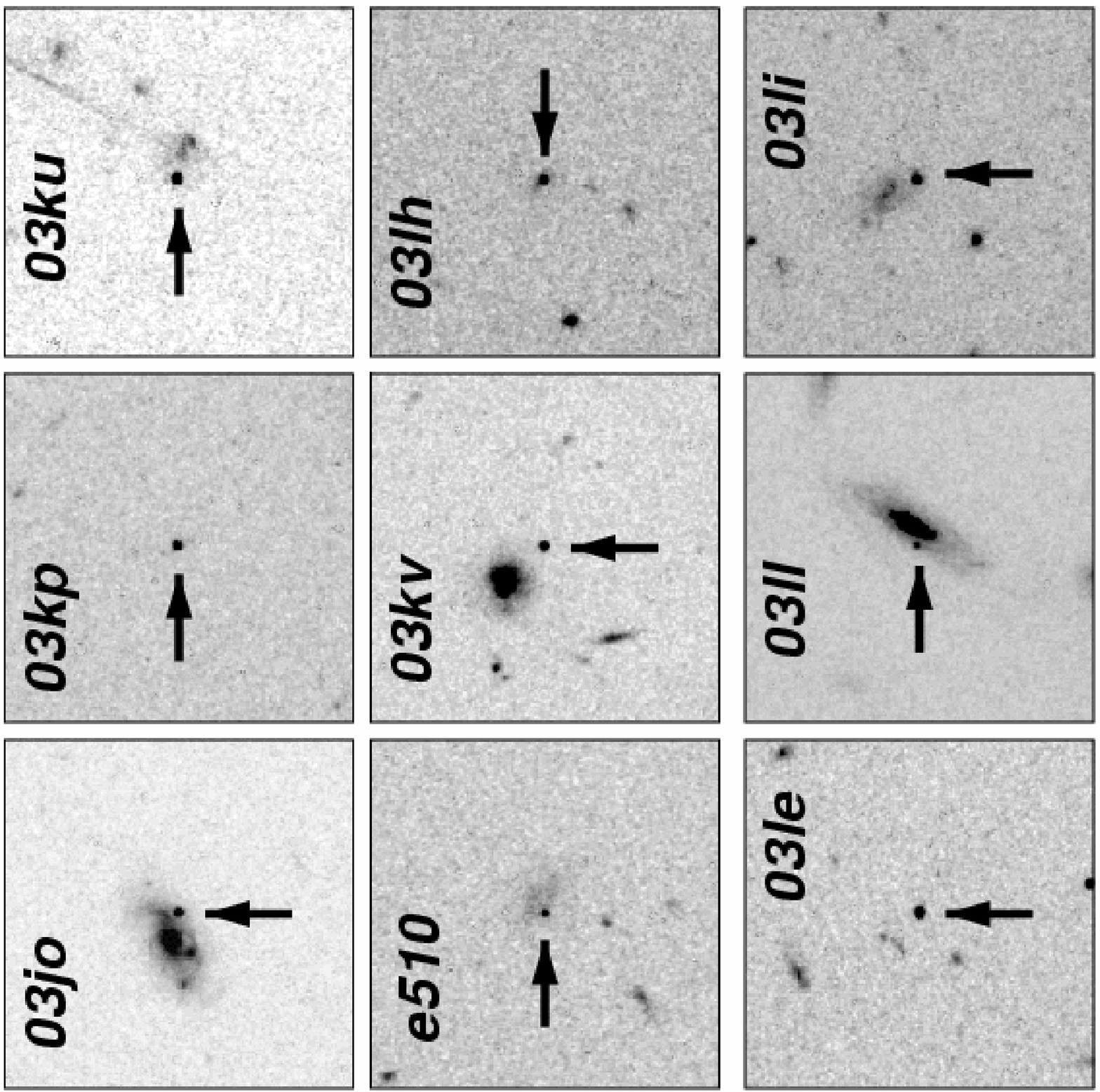] {$10'' \times 10''$ images of the nine
ESSENCE SNe discussed in this paper.  Each was taken through the
F775W filter of $HST$/ACS.  The total integration times were
3700 s (e510), 5100 s (SN 2003jo), and 4400 s (the other seven SNe).
\label{stamps}
}

\figcaption[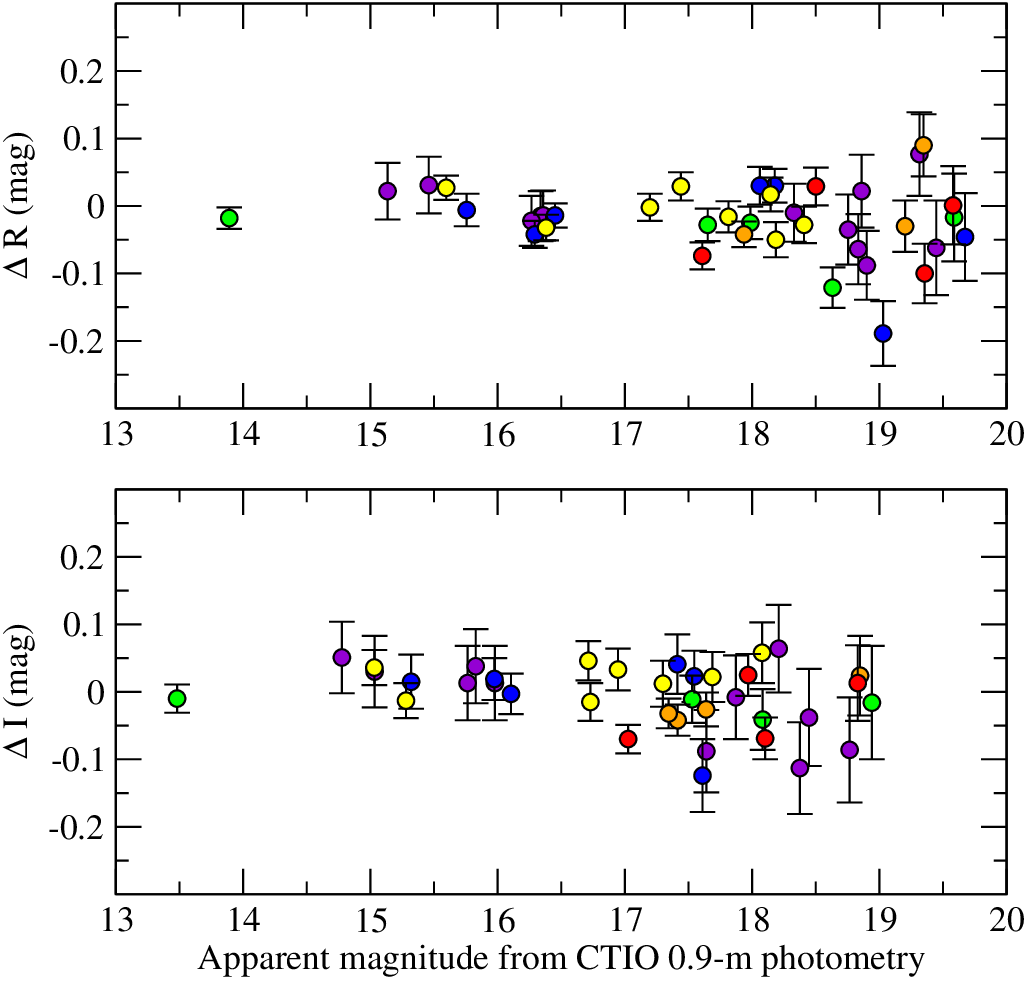] {Difference of observed $R$ and $I$ photometry
of stars near six of our SNe compared to $R$ and $I$ magnitudes derived
from SDSS data using the transformations of \citet{ASmi_etal02}.
$\Delta$ is in the sense ``observed values from CTIO 0.9-m
photometry directly tied to \citet{Lan92} standards'' {\em minus}
``values derived from SDSS photometry.''  (In the color version of this
plot different colors correspond to different ESSENCE fields.)
\label{RI_test}
}

\figcaption[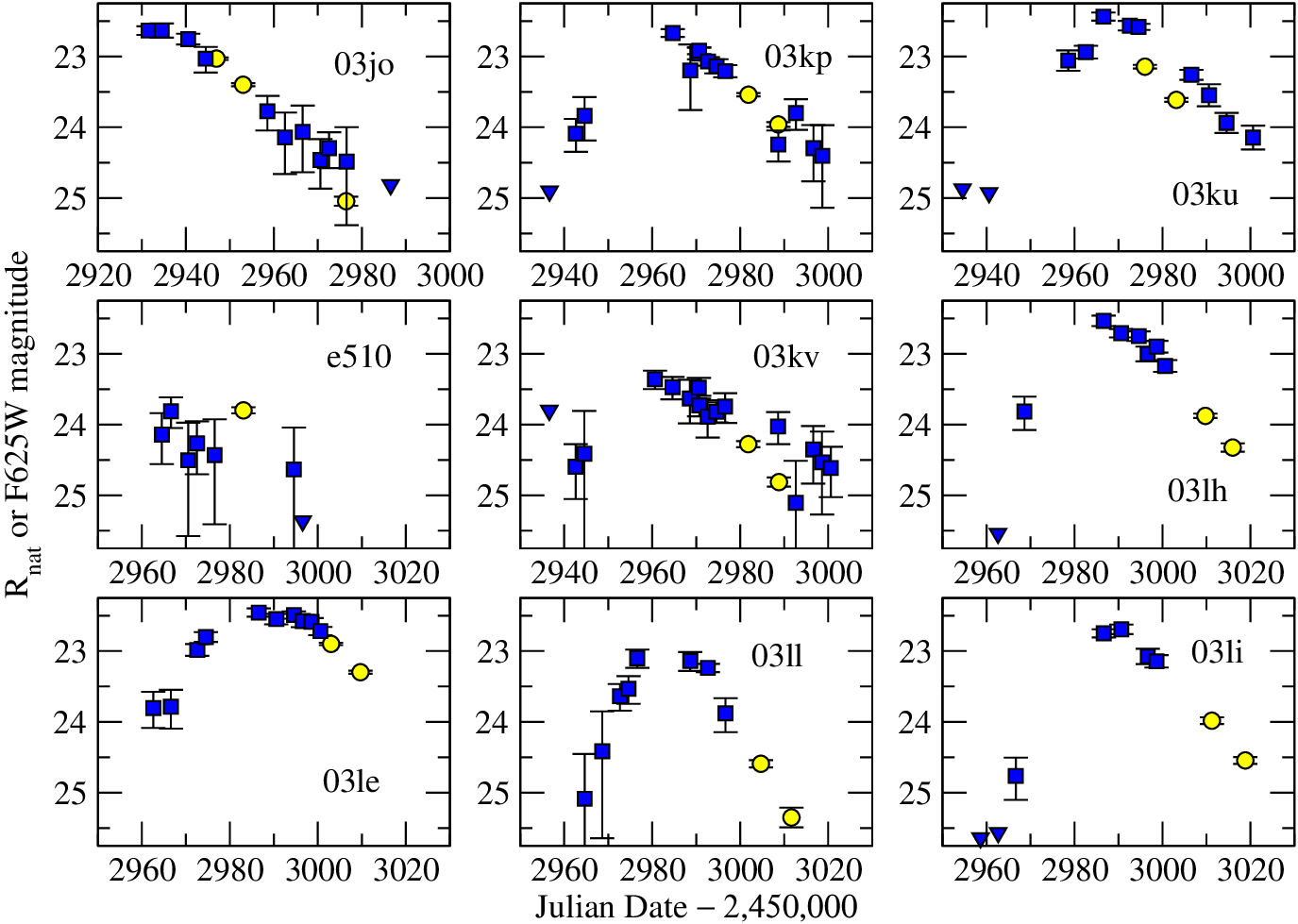] {Observed-frame natural system photometry of the nine SNe
discussed in this paper. Symbols: $R$ band = (blue) squares; $HST$/ACS F625W 
filter = (yellow) dots.  Downward pointing triangles are 1$\sigma$ upper limits 
in $R$. 
\label{R_light_curves}
}

\figcaption[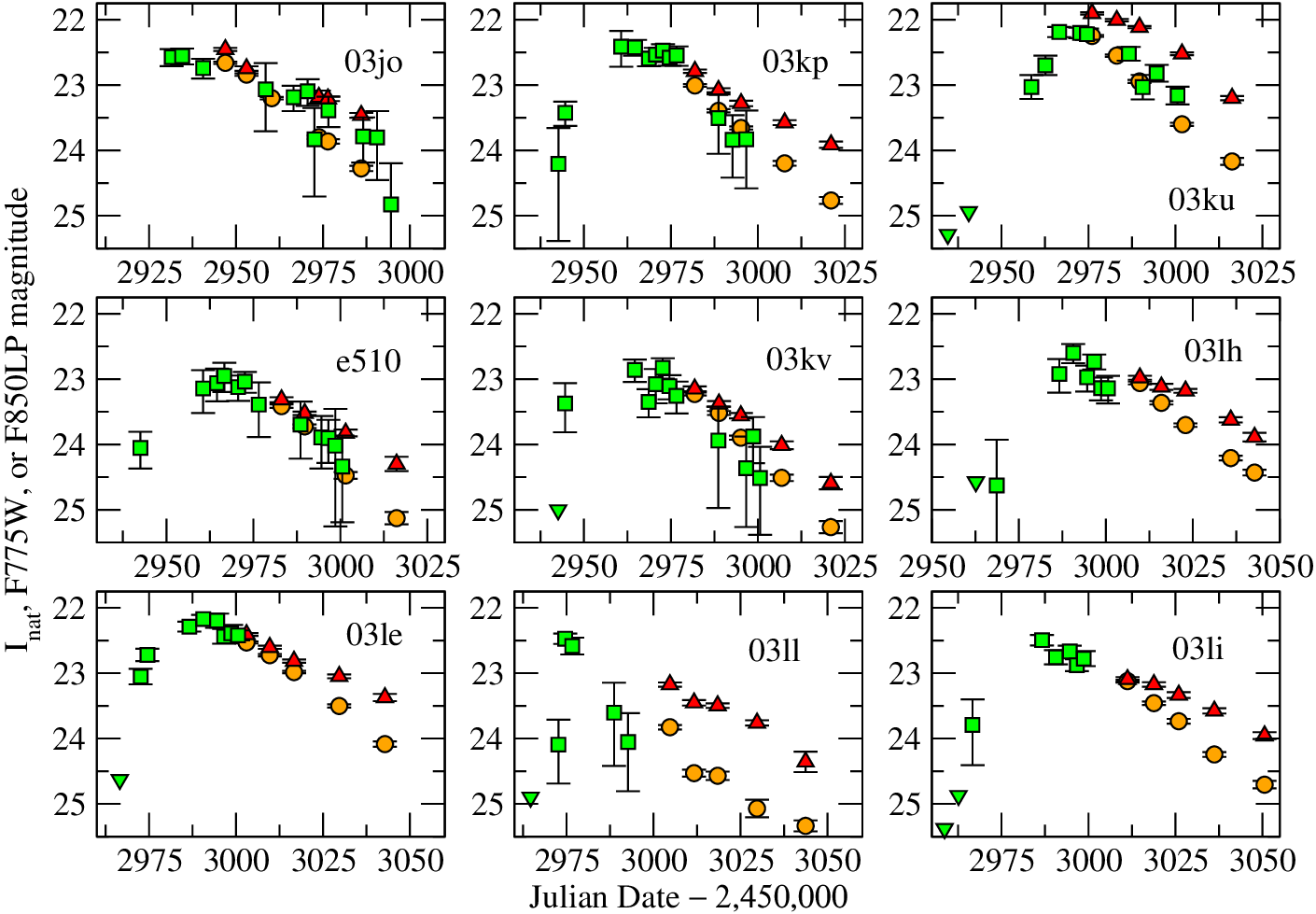] {Observed-frame natural system photometry of the nine SNe
discussed in this paper. Symbols: $I$ band = (green) squares; $HST$/ACS F775W 
filter = (orange) dots; $HST$/ACS F850LP filter = (red) upward-pointing 
triangles. Downward-pointing triangles are 1$\sigma$ upper limits in $I$.
\label{I_light_curves}
}

\figcaption[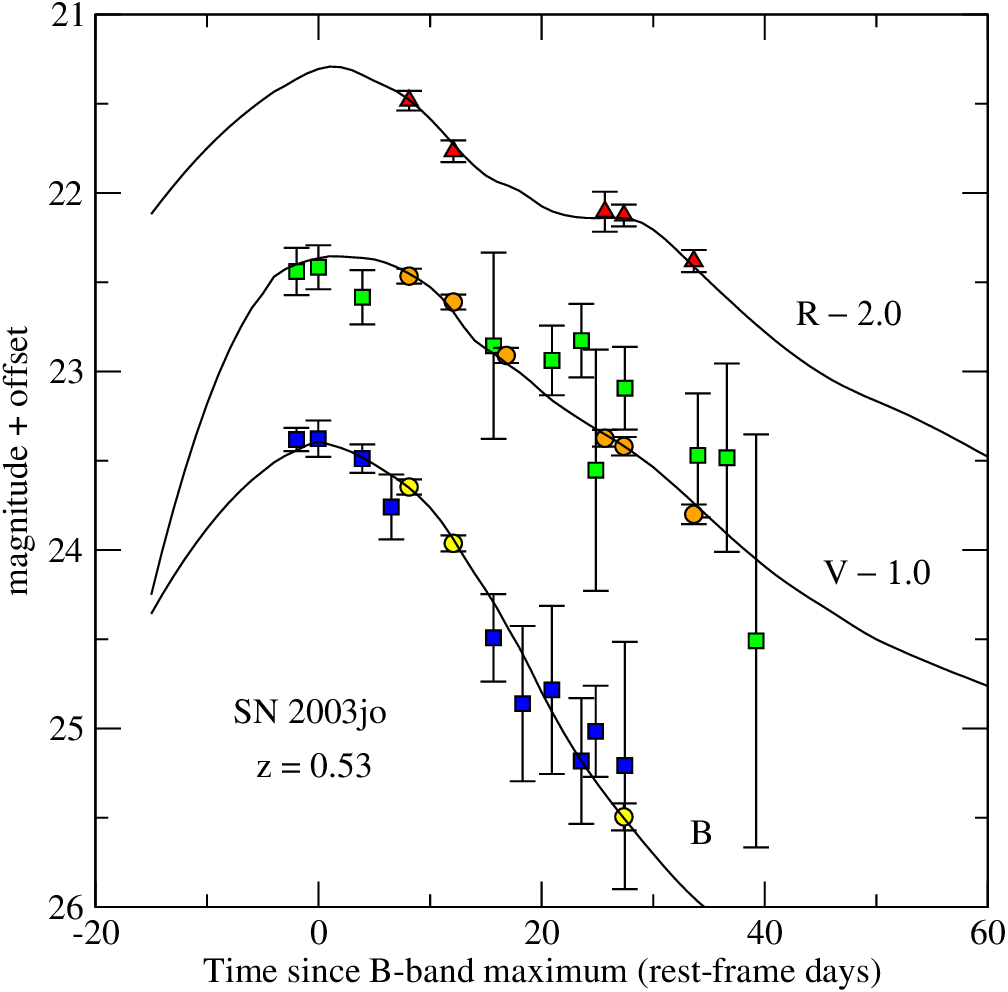,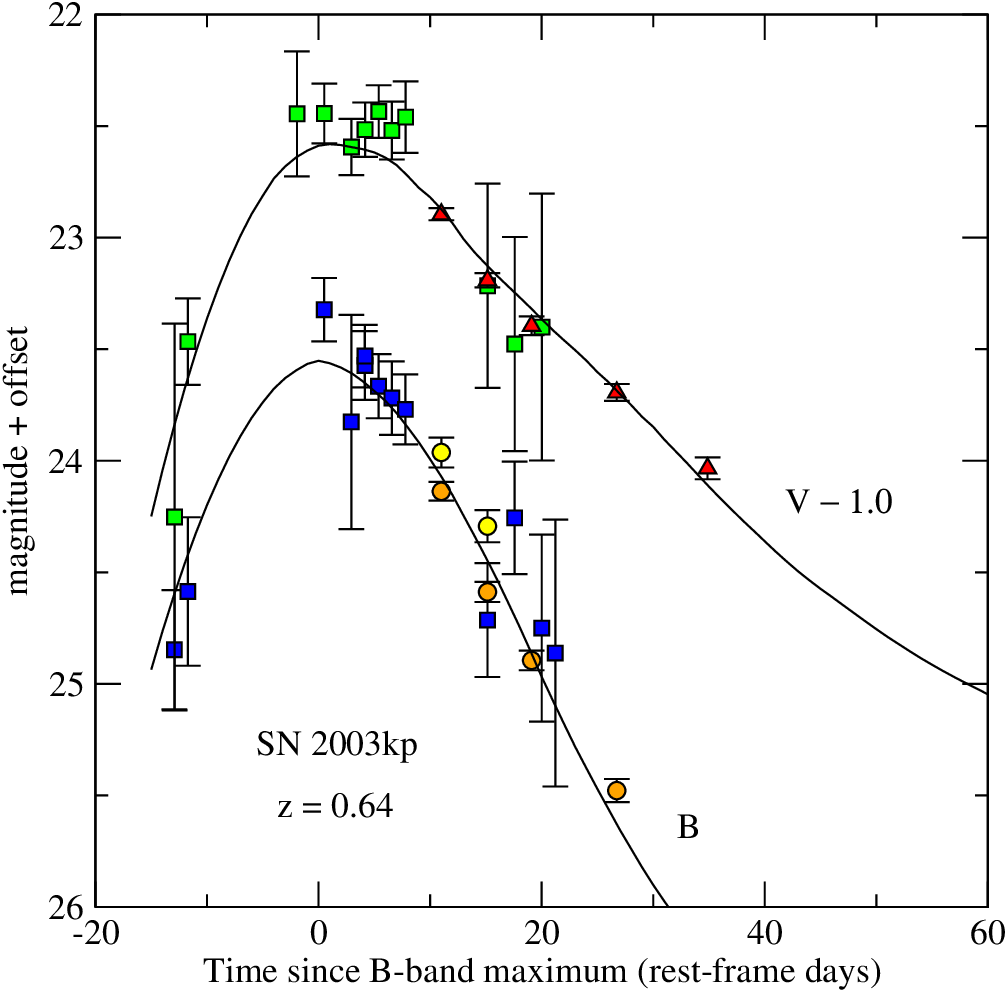,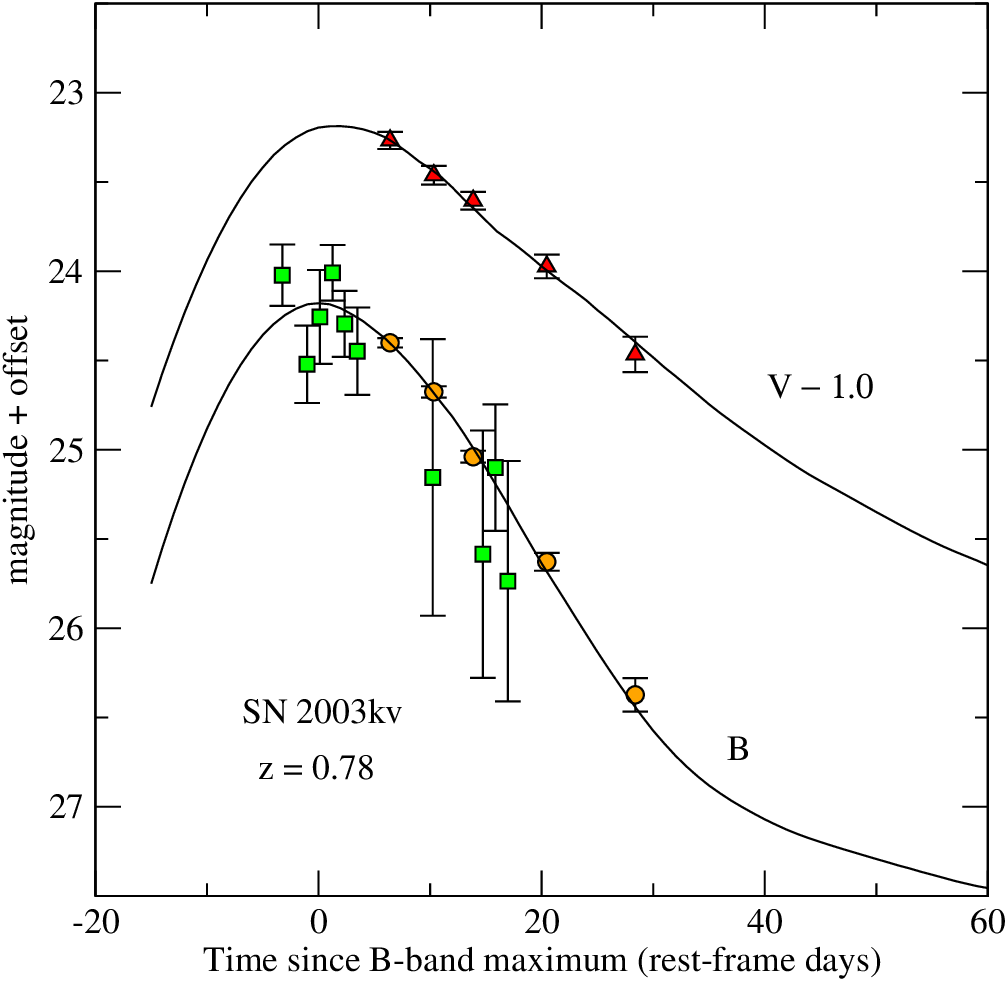,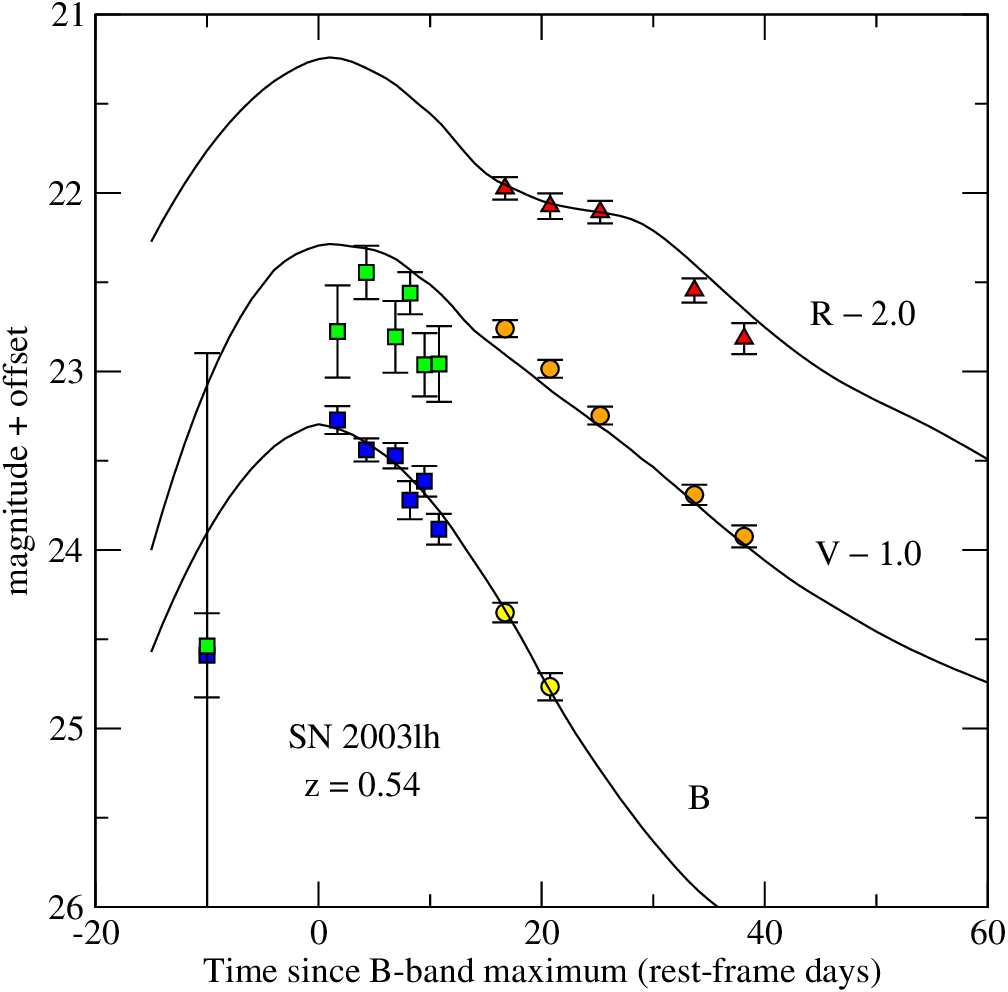,
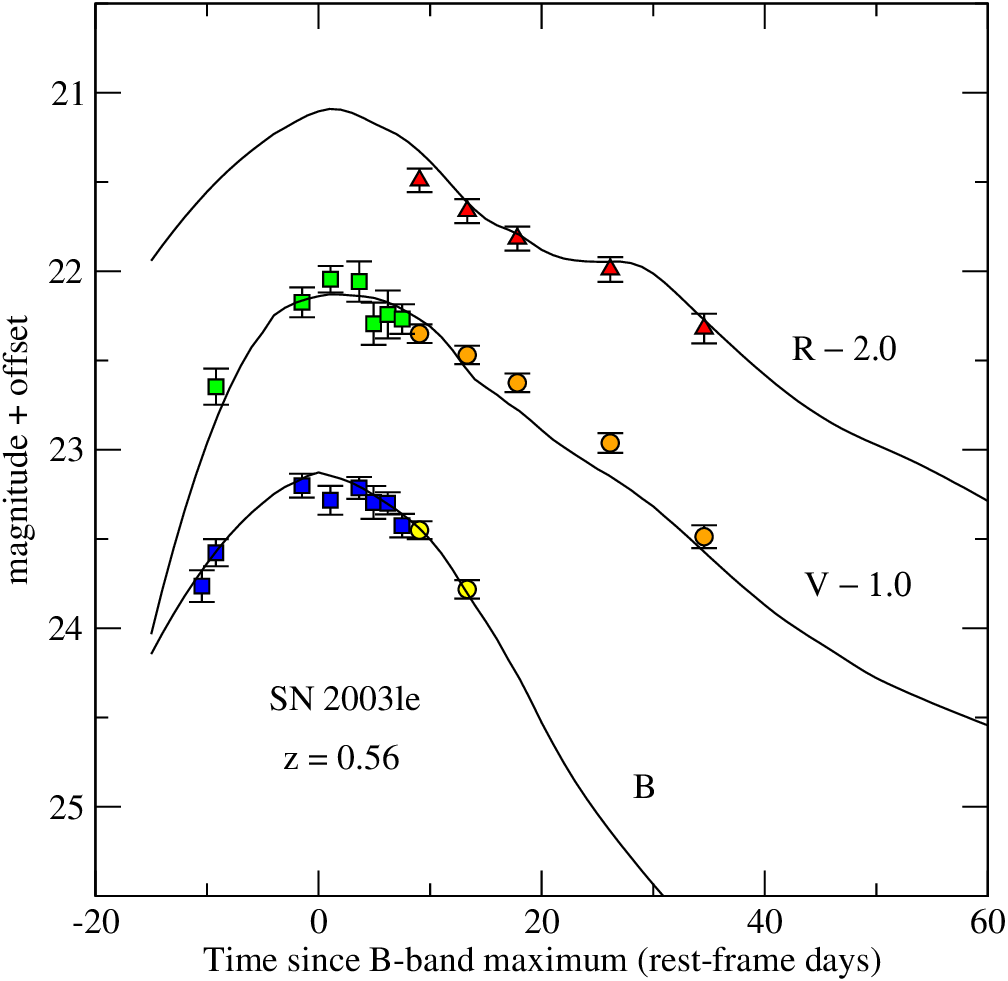,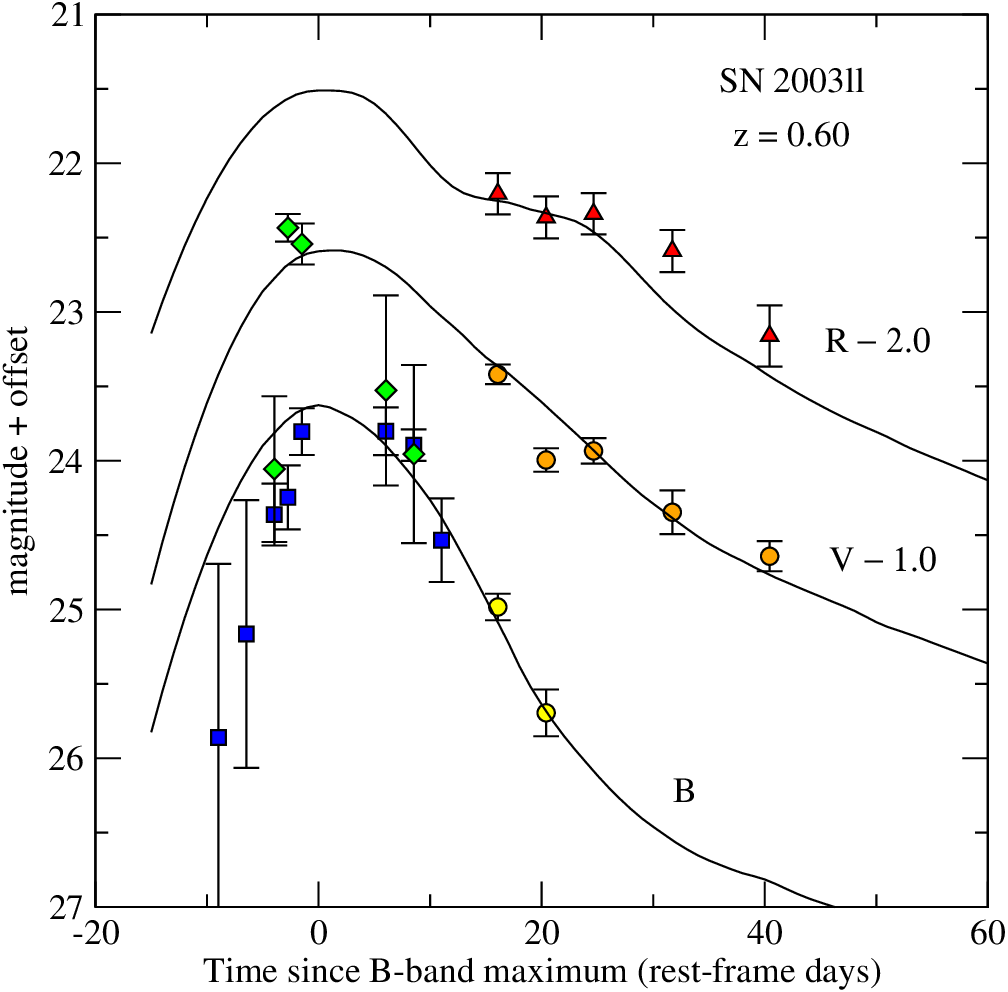,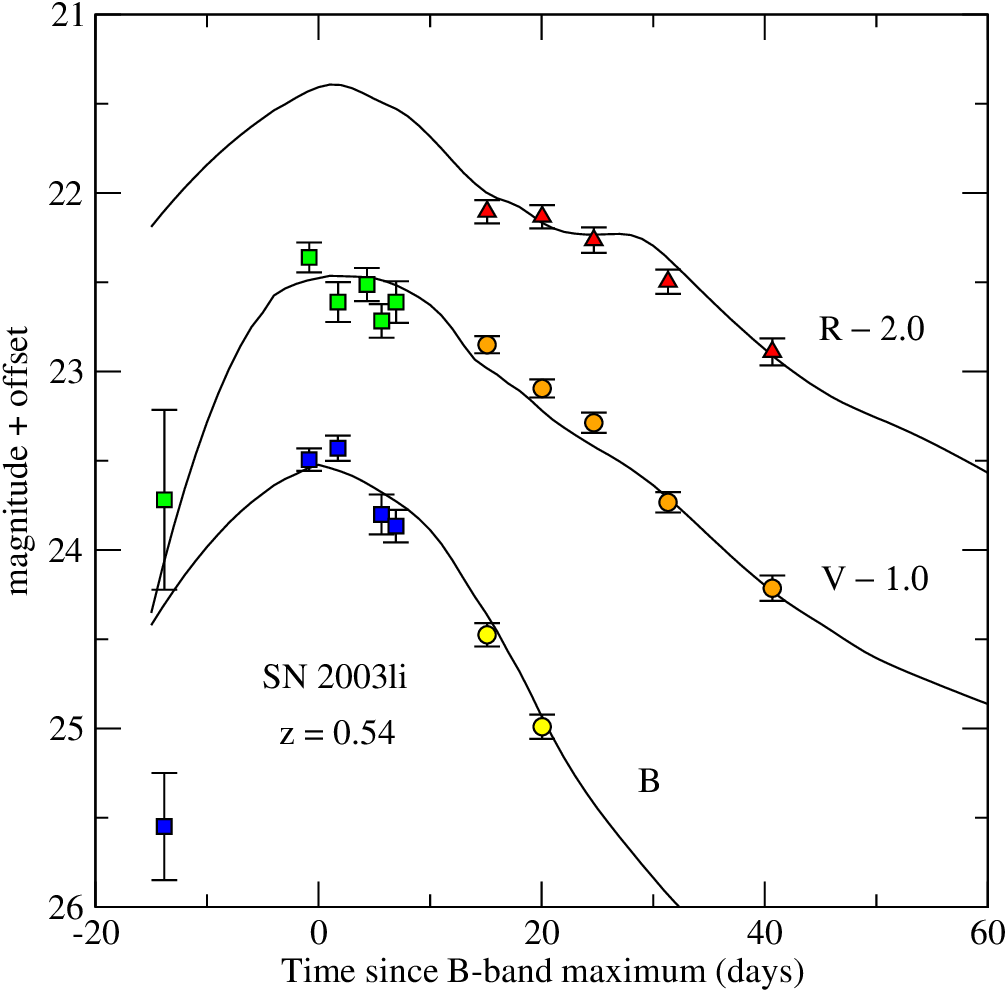] {Light-curve fits in the rest-frame
bands.  All these fits used the \dmm\ method of \citet{Pri_etal05}.
The symbols are the same as in Figures \ref{R_light_curves} 
and \ref{I_light_curves}, with one exception.  In Fig. \ref{fits}f
the diamond-shaped symbols correspond to photometry of SN~2003 ll
originally obtained in the $I$-band.
\label{fits}
}

\figcaption[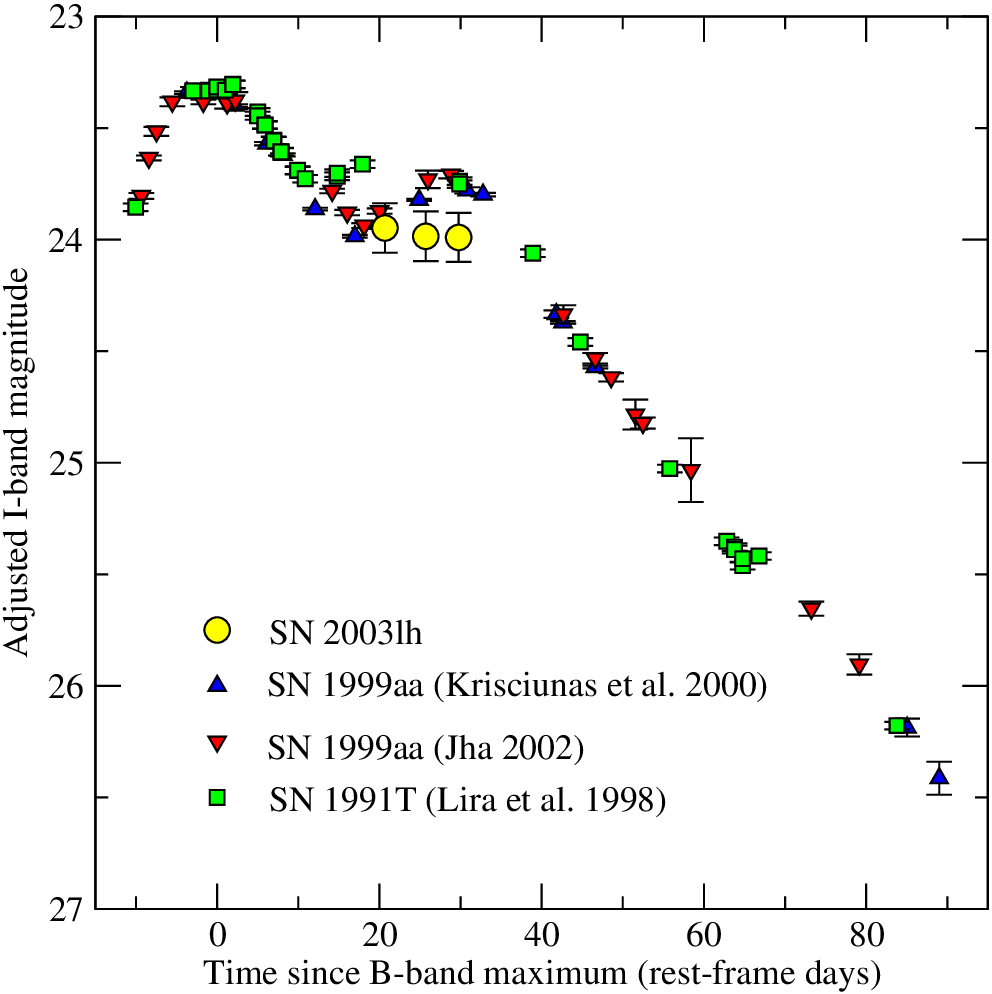] {K-corrected, extinction-corrected $I$-band data
of SN~2003lh along with $I$-band data of the slow decliners 
SN~1999aa \citep{Kri_etal00,Jha02} and SN~1991T \citep{Lir_etal98}, 
adjusted in magnitude space to the brightness of SN~2003lh
using the \dmm\ solution in Table \ref{lc_fits} and using $M_I$(max) =
$-$19.32 mag (on an $H_0$ = 65 km s$^{-1}$ Mpc$^{-1}$ scale) 
for the slowest decliners studied by  \citet{Nob_etal05}. 
\label{f011_iband}
}

\figcaption[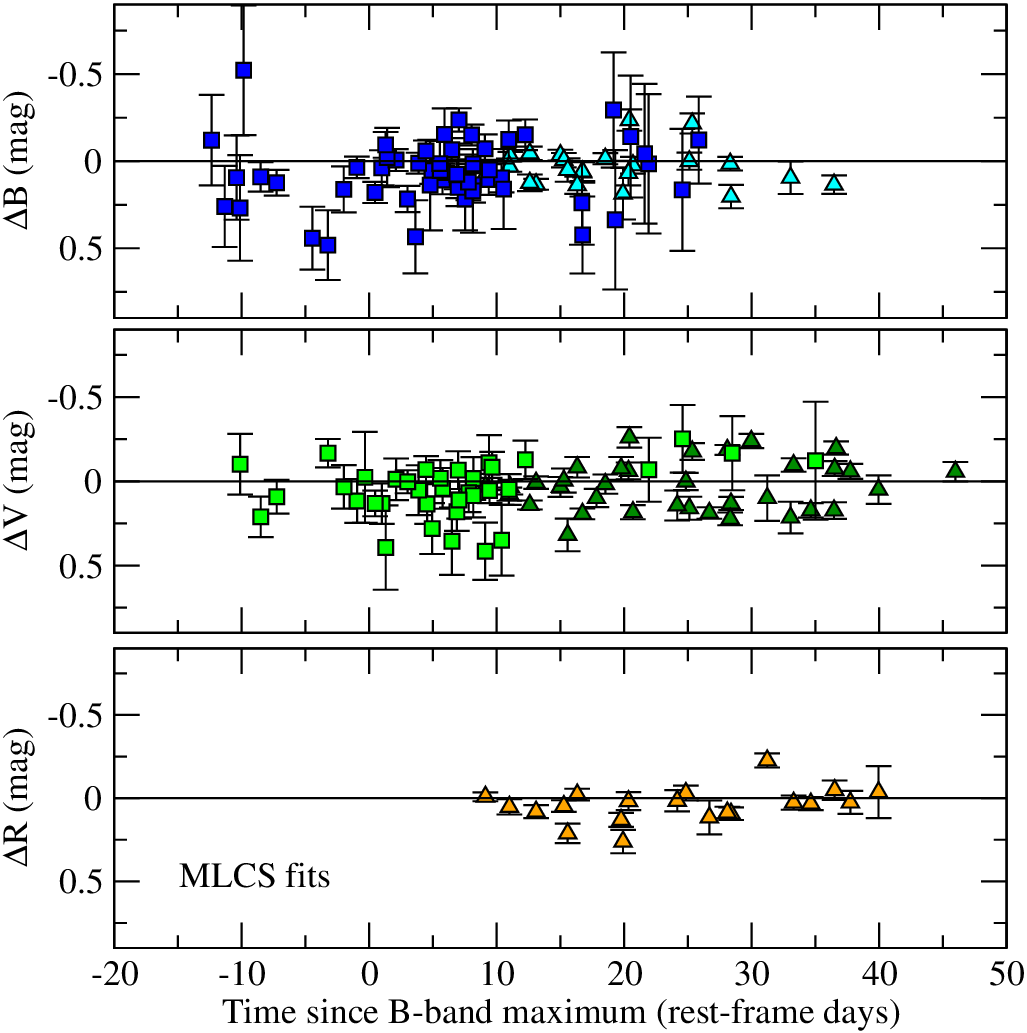] {Residuals of MLCS2k2
light-curve fits.  Here we assume no limits on the possible
values of the MLCS parameter $\Delta$.
The CTIO 4-m data are represented by squares, while
the $HST$ data are represented by triangles.
Differential magnitude here is in the sense of ``rest-frame
extinction-corrected, K-corrected data value'' {\em minus} 
``interpolated light-curve fit value.''
\label{mlcs_resids}
}

\figcaption[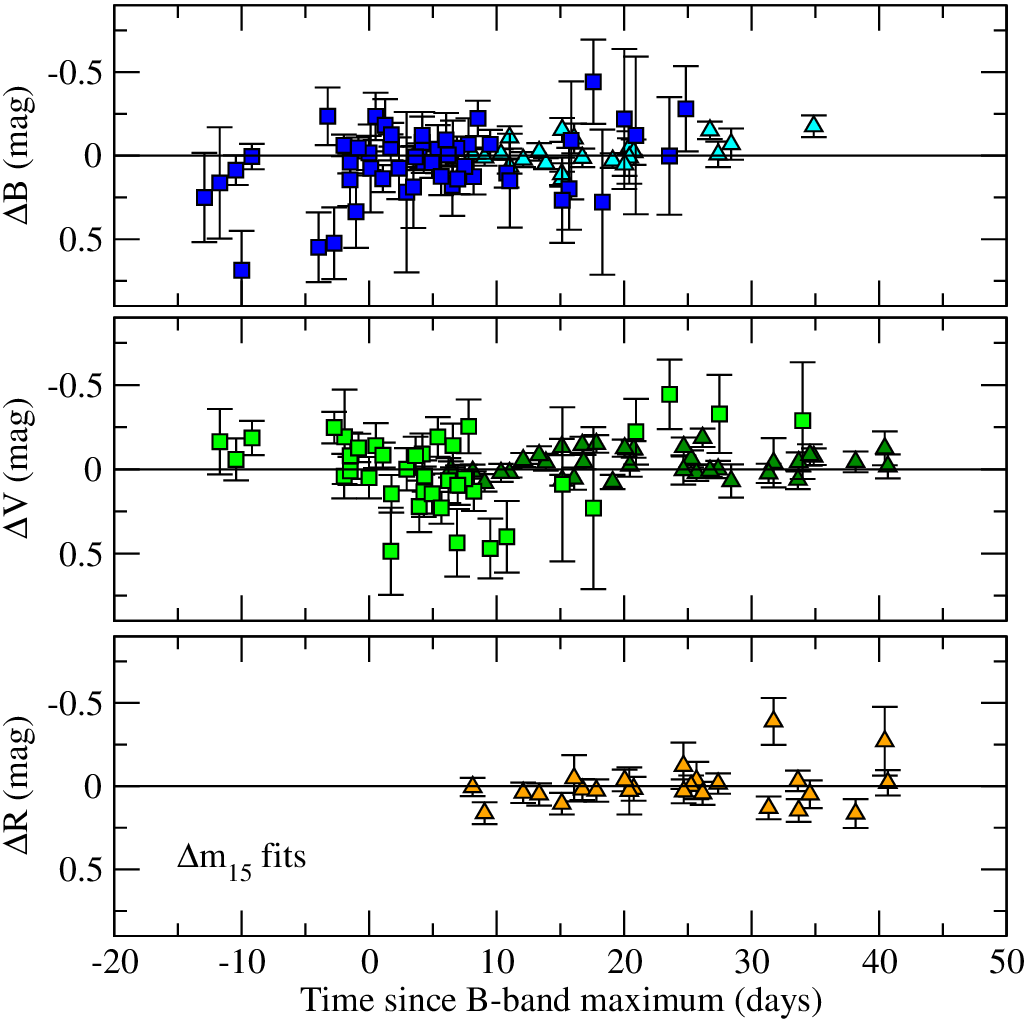] {Same as for Figure \ref{mlcs_resids}, but
residuals of data and \dmm\ light-curve fits.
\label{dm15_resids}
}

\figcaption[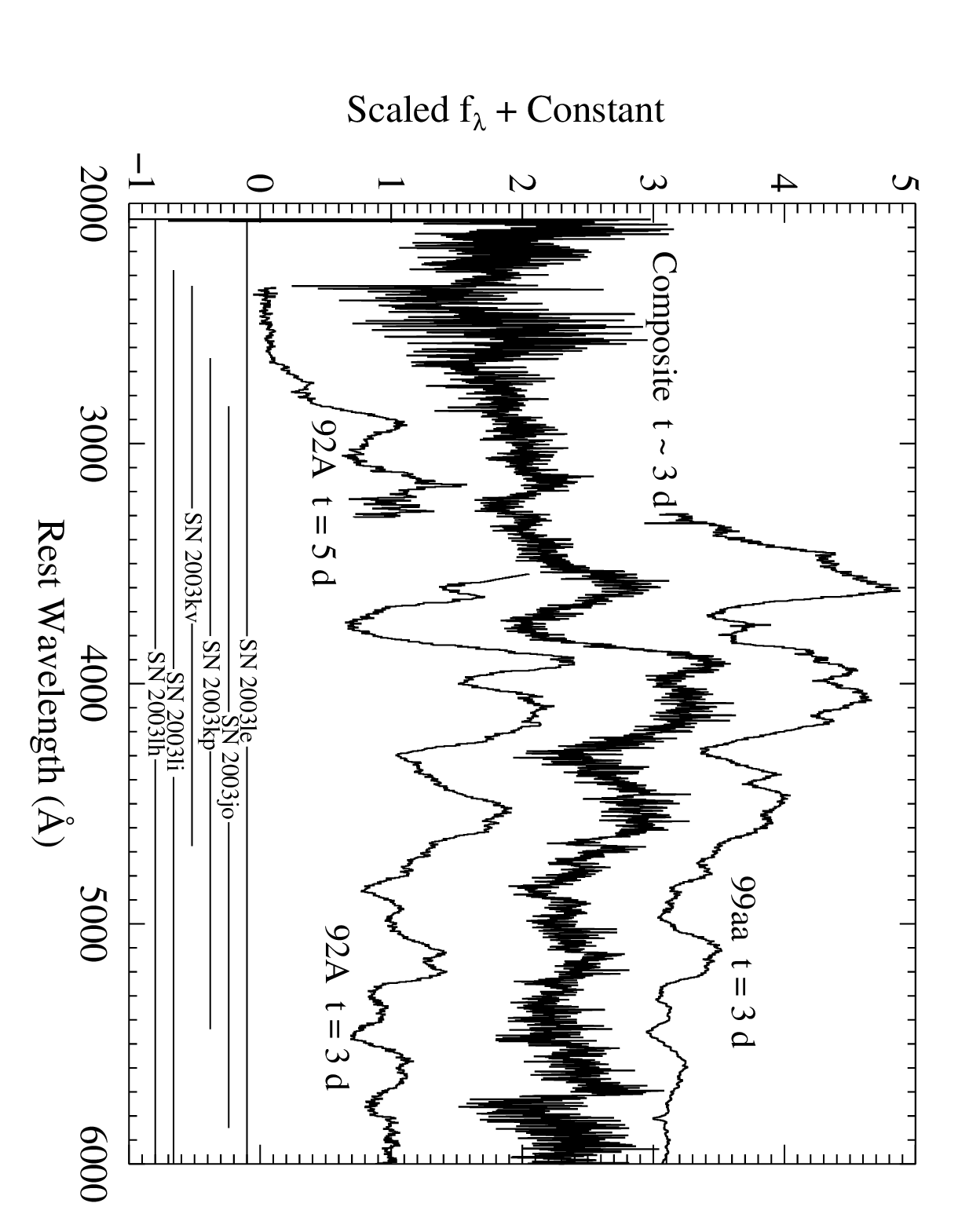] {
Composite spectrum of the six slowly declining ESSENCE objects.  SNe~1992A
and 1999aa (at ages of 3 and 5 d past maximum brightness for SN~1992A and
3 d past maximum brightness for SN~1999aa) are plotted for comparison.  
The bars at the bottom of the plot show the wavelength range of each object's
spectrum used for the composite spectrum.
\label{composite}
}

\figcaption[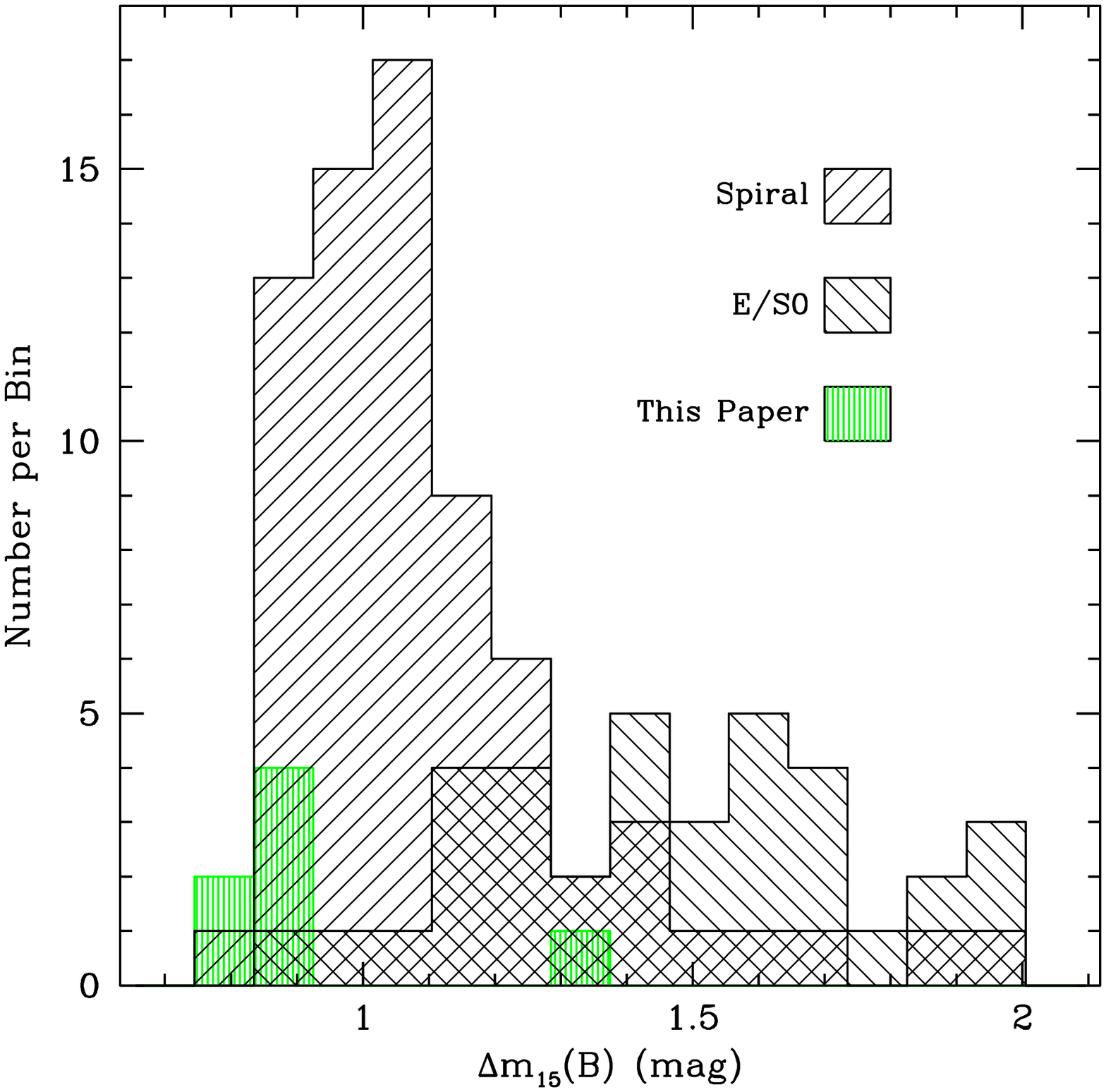] {Histogram of decline-rate values [\dmm] for
107 nearby SNe~Ia \citep{Gal_etal05} and values for seven ESSENCE SNe from 
Table \ref{lc_fits}.  Clearly, most of the ESSENCE objects have 
slow decline rates compared to the local sample.
\label{dm15_hist}
}

\figcaption[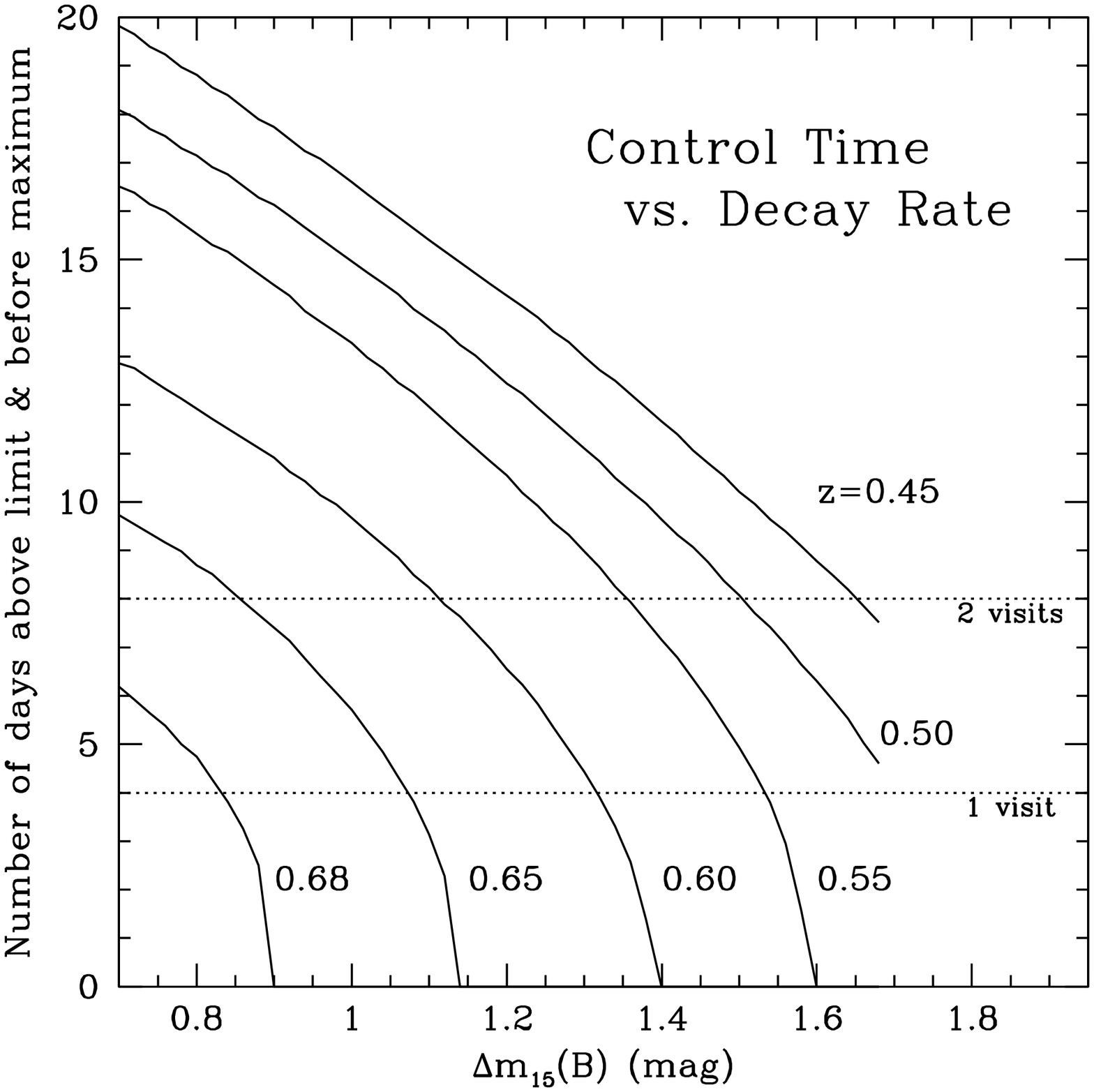] {Typical pre-maximum detection window
for SNe~Ia discovered with the CTIO 4-m telescope.  We assume
$0.9''$ seeing, S/N $\gtrsim 10$, and
a detection threshold of $R$ = 23.0 mag.  ESSENCE fields are
typically imaged (or ``visited'') every 4 days.  Under typical
seeing conditions, most of the SNe~Ia discovered with
$z \gtrsim$ 0.60 will have slow decline rates. 
\label{control_time}
}

\figcaption[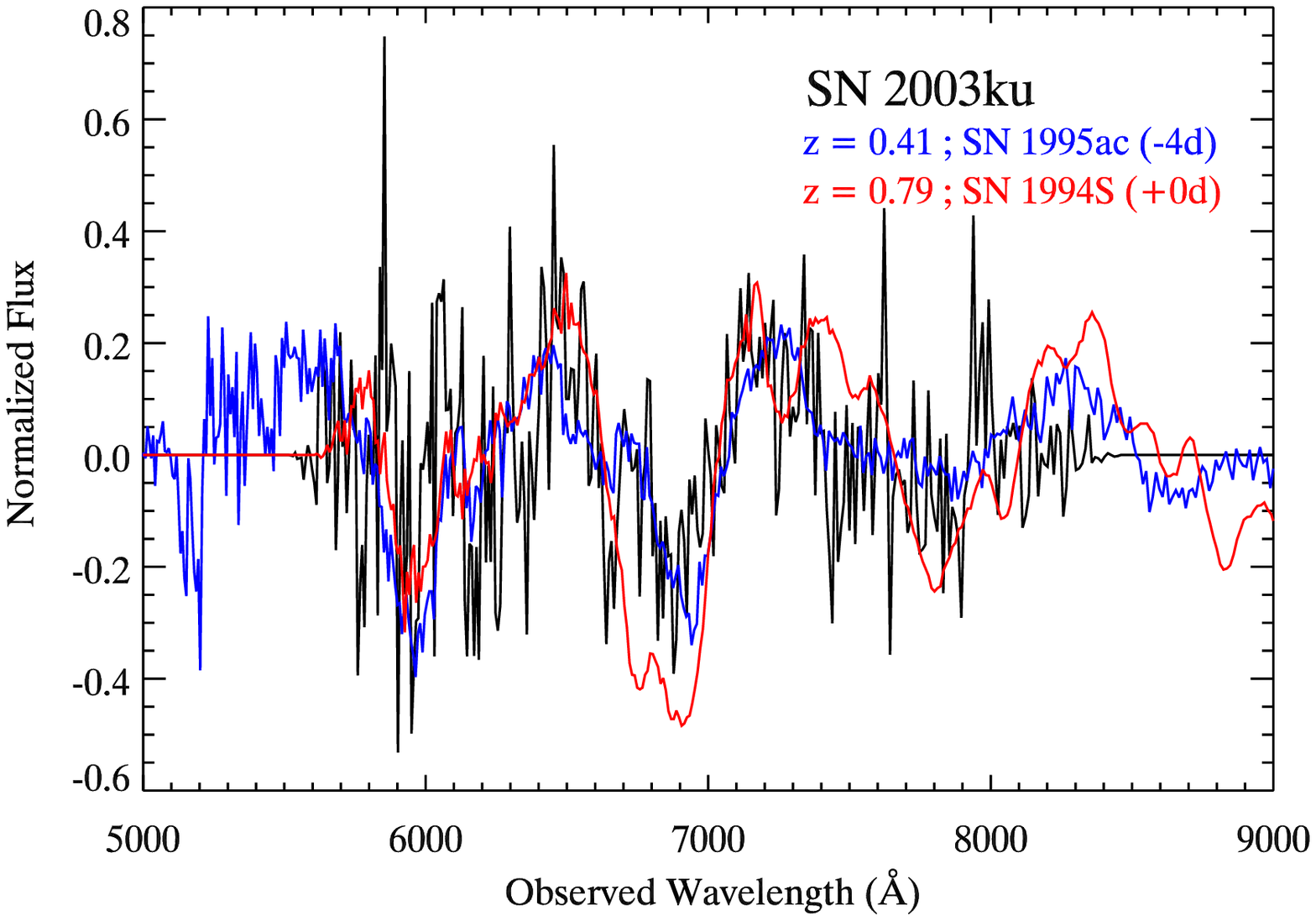] {Spectra of two nearby SNe~Ia, redshifted by the 
indicated amounts, and superimposed on the spectrum of SN~2003ku.  
``Normalized flux'' means that a pseudo-continuum has been subtracted 
from the spectra.
\label{e315_spec}
}

\figcaption[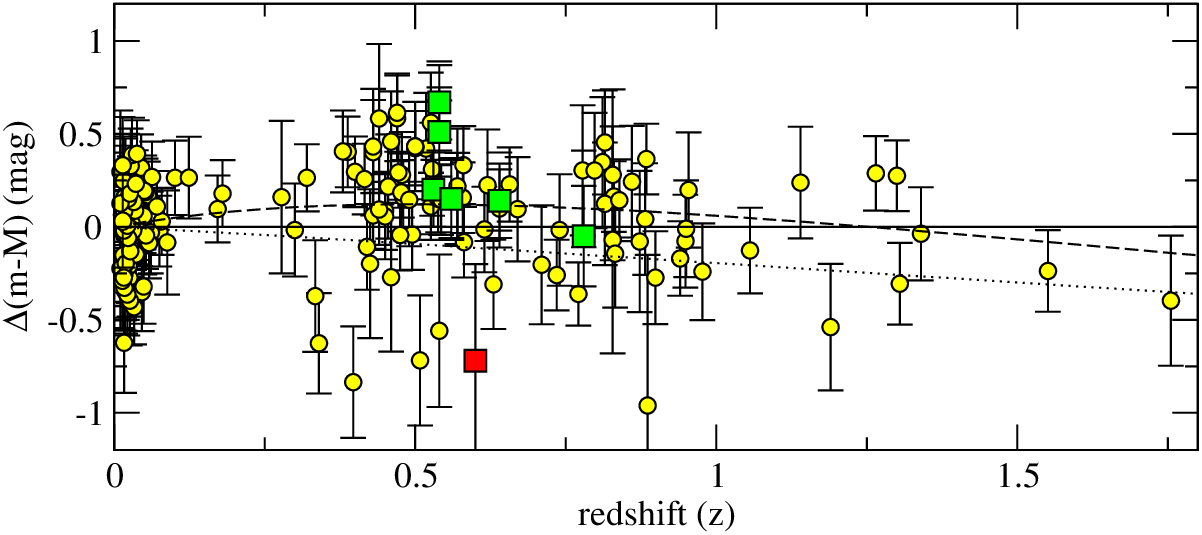] {Differential Hubble diagram for SNe~Ia.  For each object
we plot the distance modulus derived from the light curves minus the distance modulus
in an empty universe vs. the redshift.  See Table \ref{mlcs_fits}.  
The yellow dots are the ``gold'' data set of 157 SNe~Ia from \citet{Rie_etal04}.  
The ESSENCE SNe~Ia are represented by larger squares.  SN~2003ll
is represented by a red square having a distance modulus 0.72 mag {\em 
brighter} than the empty-universe model.  The dashed line is the
concordance model ($\Omega_M = 0.3, \Omega_{\Lambda} = 0.7$).  The solid horizontal
line corresponds to distance moduli in the empty-universe model.  The dotted line
corresponds to an open universe with $\Omega_M$ = 0.3 and $\Omega_{\Lambda}$ = 0.0.
\label{diff_hub}
}

\figcaption[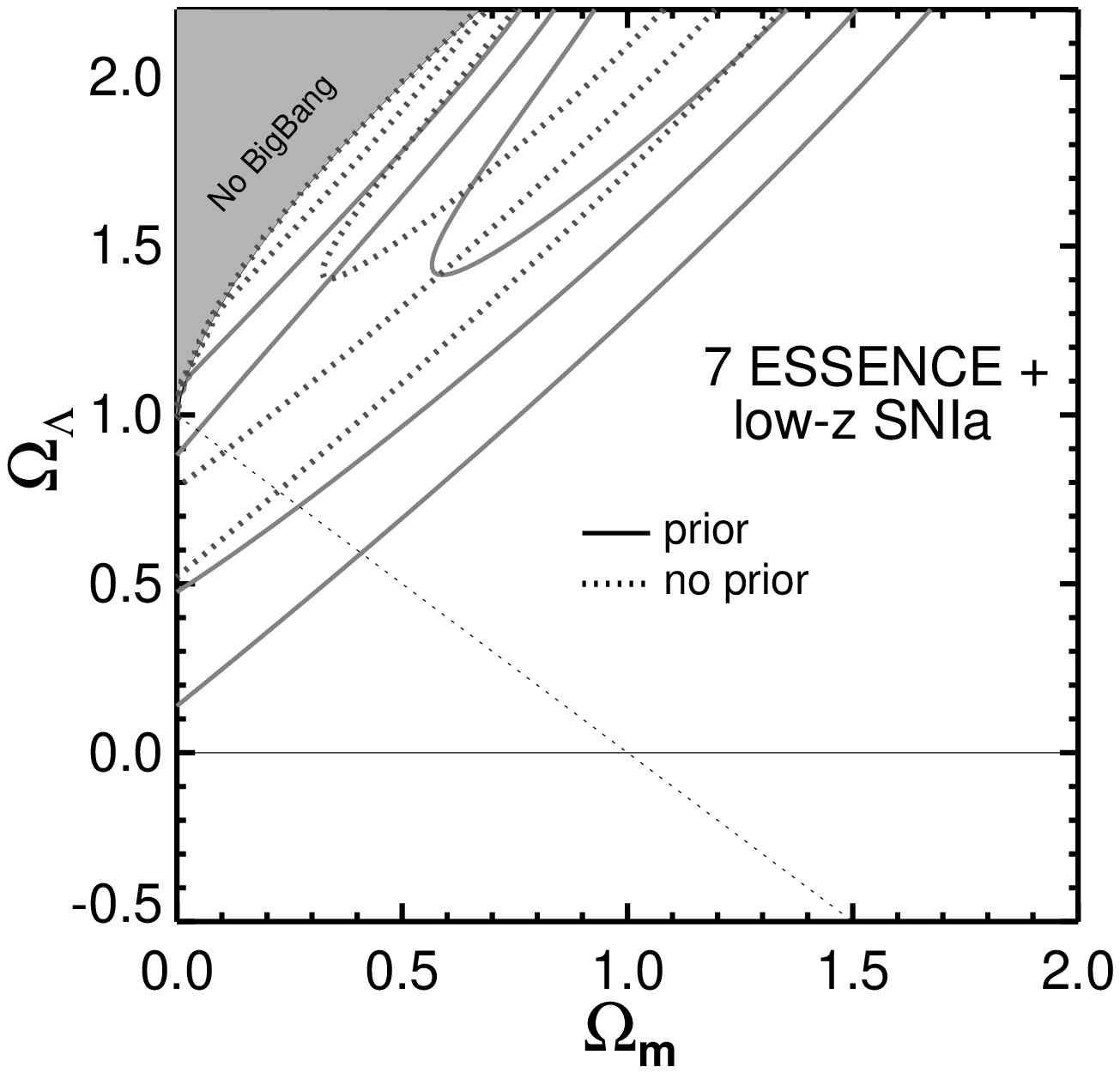] {Constraints on the 
dark energy ($\Omega_{\Lambda}$) using the nearby SNe~Ia of the ``gold''
set of \citet{Rie_etal04}, plus seven ESSENCE SNe~Ia.  
We show sets of 1$\sigma$, 2$\sigma$, and 3$\sigma$ contours.
The solid lines use a constraint that MLCS $\Delta \geq -0.40$.  
The dashed contours assume no constraints on MLCS $\Delta$.
\label{contour_local_essence} 
}

\figcaption[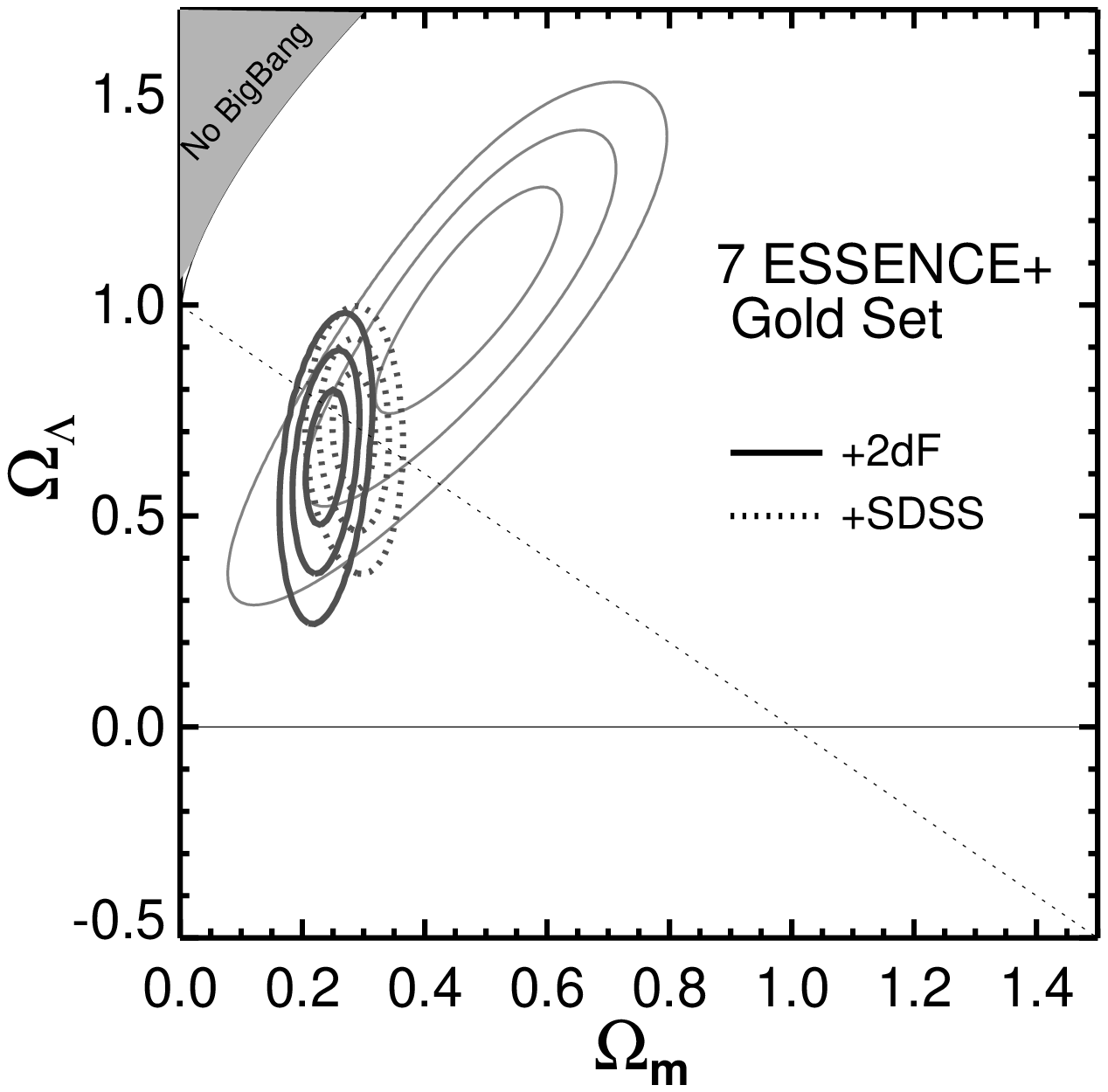] {
Constraints on the dark energy ($\Omega_{\Lambda}$) using the entire gold set of
157 SNe~Ia from \citet{Rie_etal04}, plus 7 ESSENCE SNe~Ia. 
We show sets of 1$\sigma$, 2$\sigma$, and 3$\sigma$ contours.
The thin lines come from the SN constraints only and assume the prior
constraint that MLCS $\Delta \geq -0.40$.  The thick lines include
a matter density constraint of $\Omega_M$ = 0.233 $\pm$ 0.030, in accord
with the results of \citet{Col_etal05} and H$_0$ = 72 \kms Mpc$^{-1}$ 
\citep{Fre_etal01}.  The dashed contours include a different 
matter constraint, that of \citet{Eis_etal05}, namely
$\Omega_M$ = 0.273 $\pm$ 0.025 + 0.137$\Omega_K$.  We assume $w$ = $-$1. 
\label{contour_lambda}
}

\figcaption[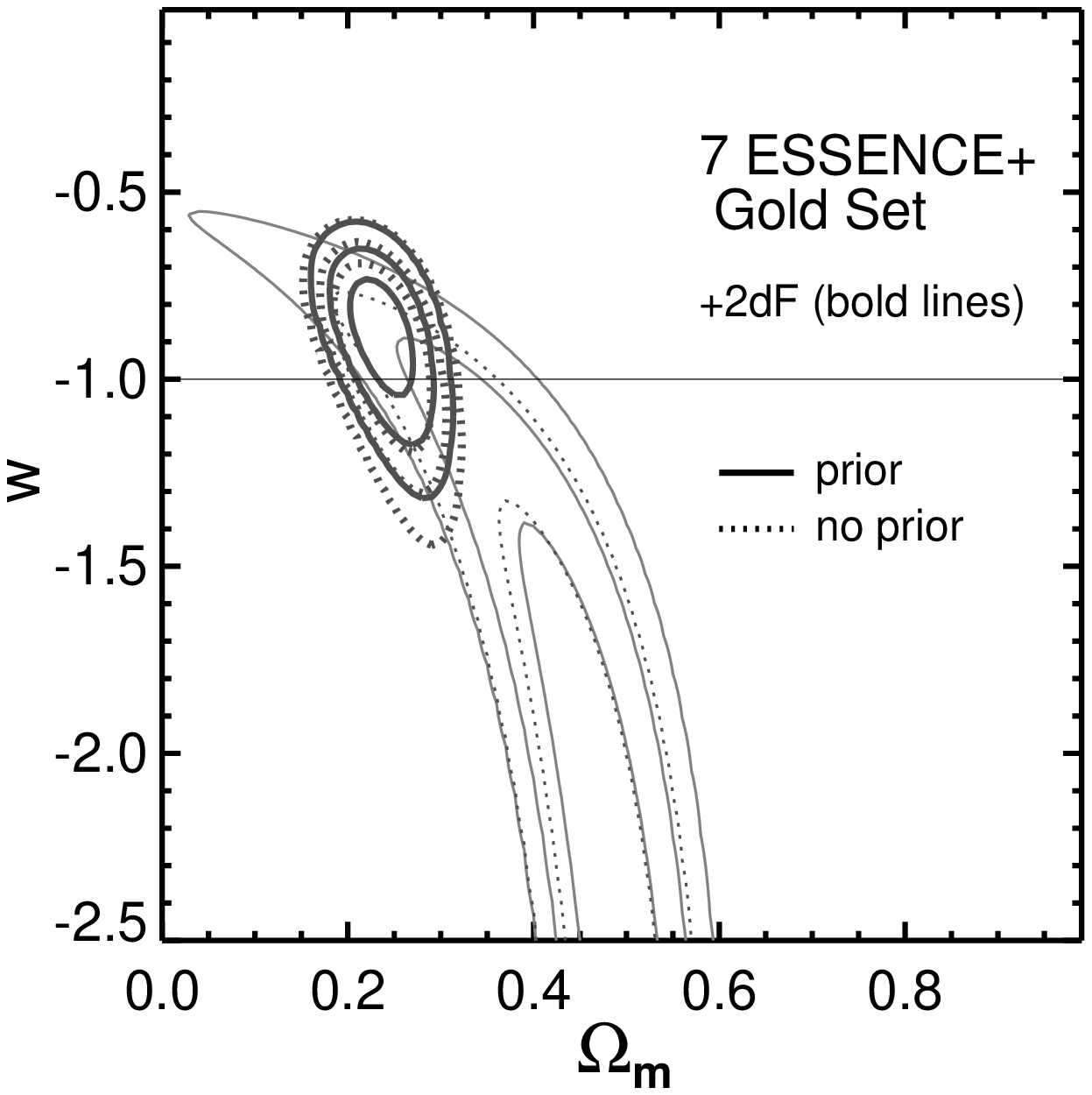] {Constraints on the equation-of-state parameter $w$
using the \citet{Rie_etal04} gold set of 157 SNe~Ia, plus 7 ESSENCE SNe~Ia.
We show sets of 1$\sigma$, 2$\sigma$, and 3$\sigma$ contours.
The thin lines come from the SN constraints only, while the thick lines are
constrained by SNe plus a matter density of $\Omega_M$ = 0.233 $\pm$ 0.030
\citep{Col_etal05}, which implies H$_0$ = 72 \kms Mpc$^{-1}$ \citep{Fre_etal01}.
The dashed contours show the effect of using no prior contraint on the 
allowable range of MLCS $\Delta$.
\label{contour_w}
}

\figcaption[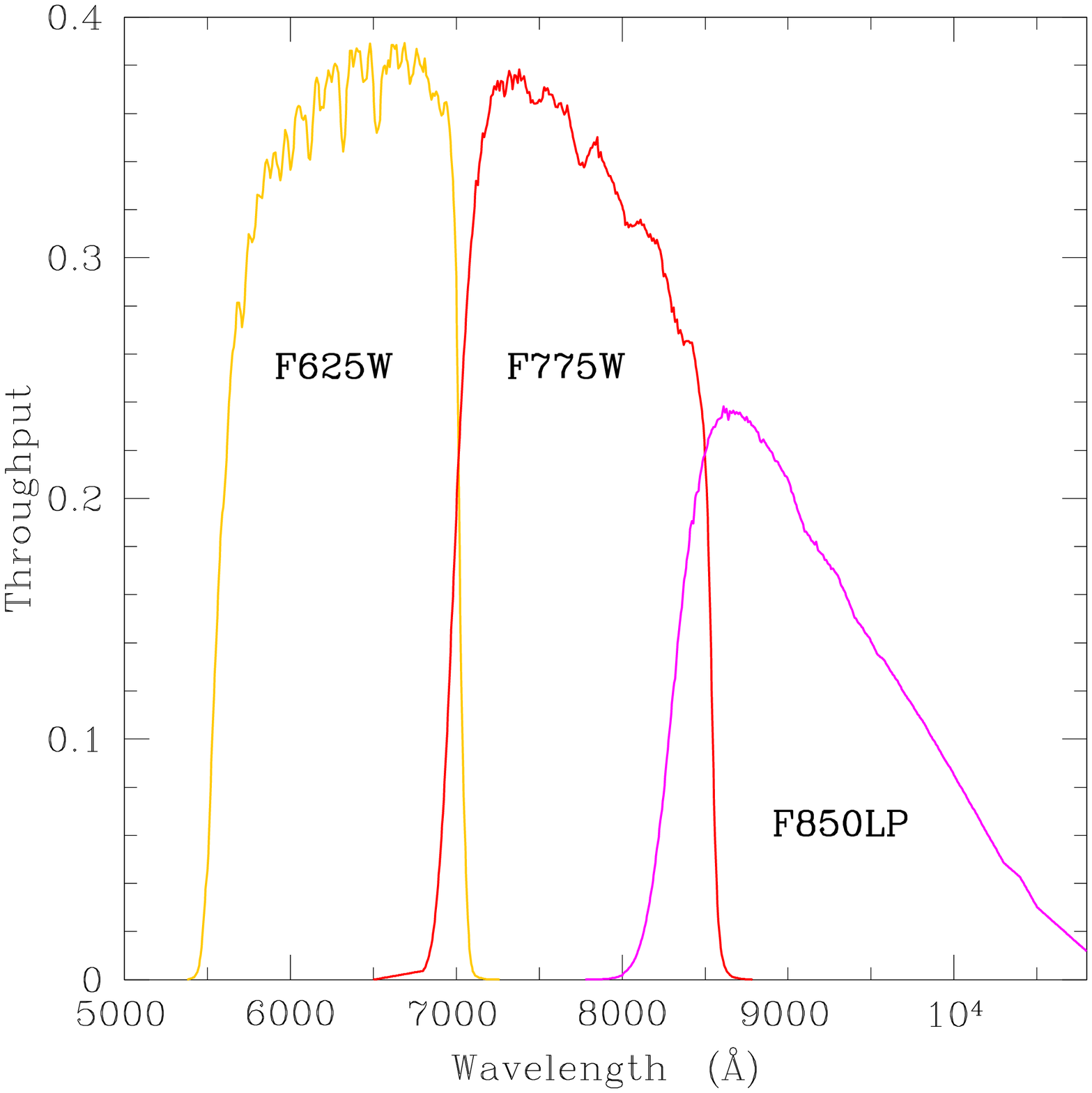] {The throughput of the F625W, F775W, and F850LP
filters as used with the $HST$/ACS.
The curves include the filter transmission functions multiplied
by the quantum efficiency as a function of wavelength.
\label{hst_filters}
}

\figcaption[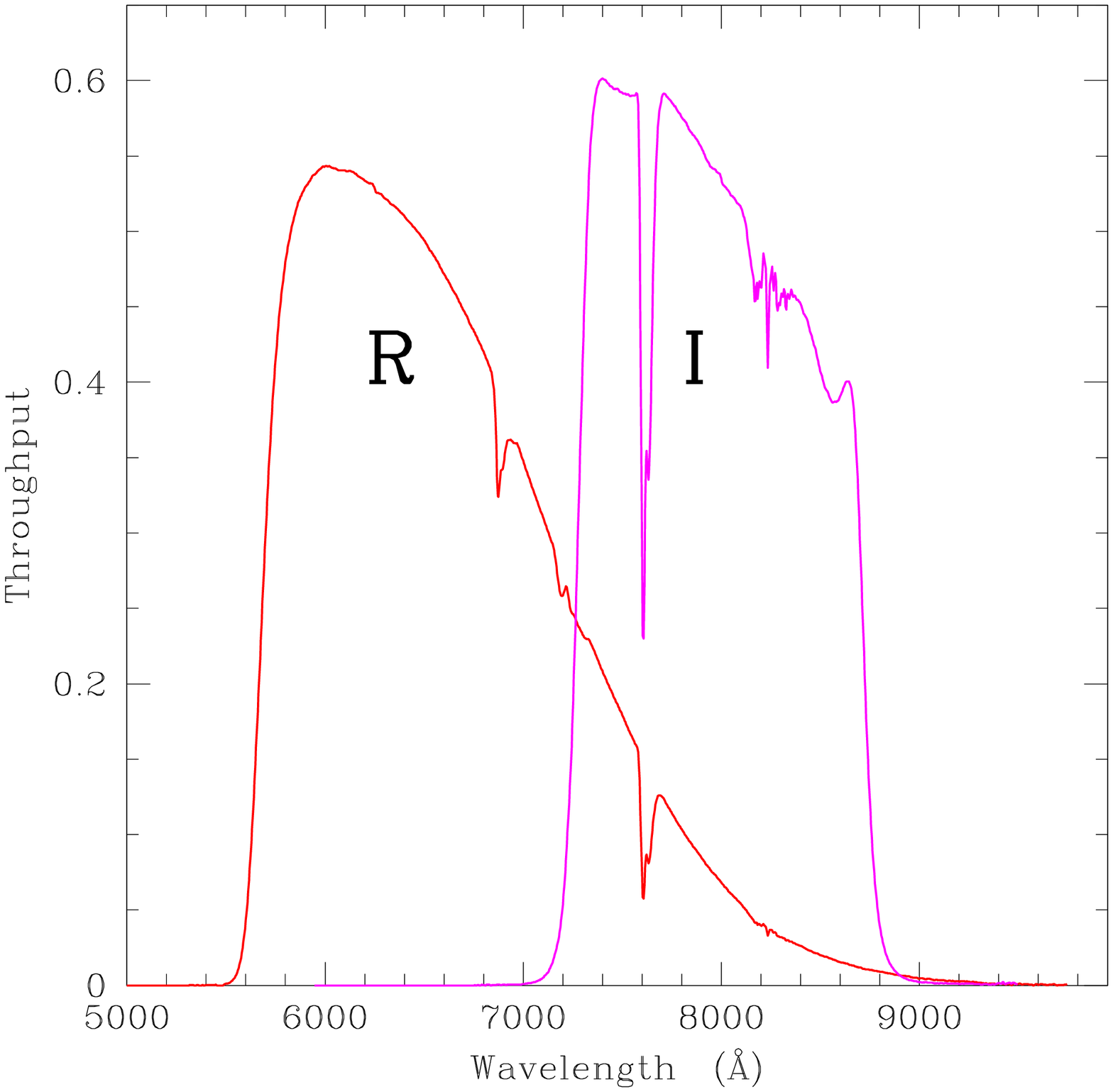] {Effective filter transmission curves of
the $R$ and $I$ filters used for ground-based photometry with
the CTIO 4-m telescope.  These curves show the fractional transmission
as a function of wavelength.  We included the filter transmission
functions determined in the laboratory, one aluminum reflection for the
effect of the primary mirror, the quantum efficiency of the CCDs, the 
atmospheric extinction, and the major telluric absorption lines.
To obtain the curves used by \citet{Bes90} one must multiply these
curves by the wavelength.
\label{CTIO4m_filters}
}

\clearpage

\begin{figure}
\plotfiddle{stamps.ps}{0.6in}{-90.}{400.}{400.}{50}{0}
{\center Krisciunas {\it et al.} Fig. \ref{stamps}}
\end{figure}

\begin{figure}
\plotone{RI_test.eps}
{\center Krisciunas {\it et al.} Fig. \ref{RI_test}}
\end{figure}

\begin{figure}
\plotone{R.eps}
{\center Krisciunas {\it et al.} Fig. \ref{R_light_curves}}
\end{figure}

\begin{figure}
\plotone{I.eps}
{\center Krisciunas {\it et al.} Fig. \ref{I_light_curves}}
\end{figure}

\begin{figure}
\plotone{d033.eps}
{\center Krisciunas {\it et al.} Fig. \ref{fits}a}
\end{figure}

\begin{figure}
\plotone{e147.eps}
{\center Krisciunas {\it et al.} Fig. \ref{fits}b}
\end{figure}

\begin{figure}
\plotone{e531.eps}
{\center Krisciunas {\it et al.} Fig. \ref{fits}c}
\end{figure}

\begin{figure}
\plotone{f011.eps}
{\center Krisciunas {\it et al.} Fig. \ref{fits}d}
\end{figure}

\begin{figure}
\plotone{f041.eps}
{\center Krisciunas {\it et al.} Fig. \ref{fits}e}
\end{figure}

\begin{figure}
\plotone{f216.eps}
{\center Krisciunas {\it et al.} Fig. \ref{fits}f}
\end{figure}

\begin{figure}
\plotone{f244.eps}
{\center Krisciunas {\it et al.} Fig. \ref{fits}g}
\end{figure}

\clearpage

\begin{figure}
\plotone{f011_i.eps}
{\center Krisciunas {\it et al.} Fig. \ref{f011_iband}}
\end{figure}

\begin{figure}
\plotone{mlcs_resids.eps}
{\center Krisciunas {\it et al.} Fig. \ref{mlcs_resids}}
\end{figure}

\begin{figure}
\plotone{dm15_resids.eps}
{\center Krisciunas {\it et al.} Fig. \ref{dm15_resids}}
\end{figure}

\begin{figure}
\plotfiddle{composite.ps}{0.6in}{90.}{500.}{500.}{-50}{0}
{\center Krisciunas {\it et al.} Fig. \ref{composite}}
\end{figure}

\begin{figure}
\plotone{dm15_hist.ps}
{\center Krisciunas {\it et al.} Fig. \ref{dm15_hist}}
\end{figure}

\begin{figure}
\plotone{control_time.ps}
{\center Krisciunas {\it et al.} Fig. \ref{control_time}}
\end{figure}

\begin{figure}
\plotone{e315_specfig.ps}
{\center Krisciunas {\it et al.} Fig. \ref{e315_spec}}
\end{figure}

\begin{figure}
\plotone{gold.eps}
{\center Krisciunas {\it et al.} Fig. \ref{diff_hub}}
\end{figure}

\begin{figure}
\plotone{contour_local_essence.ps}
{\center Krisciunas {\it et al.} Fig. \ref{contour_local_essence}}
\end{figure}

\begin{figure}
\plotone{contour_lambda.ps}
{\center Krisciunas {\it et al.} Fig. \ref{contour_lambda}}
\end{figure}

\begin{figure}
\plotone{contour_w.ps}
{\center Krisciunas {\it et al.} Fig. \ref{contour_w}}
\end{figure}

\clearpage

\begin{figure}
\plotone{hst_filters.eps}
{\center Krisciunas {\it et al.} Fig. \ref{hst_filters}}
\end{figure}

\begin{figure}
\plotone{CTIO4m_filters.ps}
{\center Krisciunas {\it et al.} Fig. \ref{CTIO4m_filters}}
\end{figure}

\end{document}